\documentclass[a4paper,fleqn]{cas-dc}

\usepackage[]{subfig}
\usepackage[numbers]{natbib}
\usepackage[]{hyperref}
\usepackage{soul, color, xcolor}
\soulregister\cite7
\soulregister\ref7 

\def\tsc#1{\csdef{#1}{\textsc{\lowercase{#1}}\xspace}}
\tsc{WGM}
\tsc{QE}

\begin{document}
\let\WriteBookmarks\relax
\def\floatpagepagefraction{1}
\def\textpagefraction{.001}

\shorttitle{EEGDiR: Electroencephalogram denoising network}    

\shortauthors{short author list for running head}  

\title [mode = title]{EEGDiR: Electroencephalogram denoising network for temporal information storage and global modeling through Retentive Network}  



%

\author[1]{Bin Wang}

\fnmark[1]

\ead{woldier@foxmail.com}


\credit{Conceptualization of this study, Methodology, Software}

\author[1]{Fei Deng}
\cormark[1]
\fnmark[1]

\ead{dengfei@cdut.edu.cn}



\author[2]{Peifan Jiang}
\fnmark[2]
\ead{jpeifan@qq.com}

\affiliation[1]{organization={College of Computer Science and Cyber Security},
            addressline={Chengdu University of Technology}, 
            city={ChengDu},
            postcode={629100}, 
            state={SiChuan},
            country={China}}

\affiliation[2]{organization={ College of Geophysics},
	addressline={Chengdu University of Technology}, 
	city={ChengDu},
	postcode={629100}, 
	state={SiChuan},
	country={China}}

\cortext[1]{Corresponding author. College of Computer Science and Cyber Security, Chengdu University of Technology, ChengDu, 629100, China}

\fntext[1]{Bin Wang and Fei Deng are with the College of Computer Science and Cyber Security, Chengdu University of Technology, Chengdu 610059, China(e-mail: woldier@foxmail.com, e-mail: dengfei@cdut.edu.cn).}
\fntext[2]{Peifan Jiang are with the College of Geophysics, Chengdu University of Technology, Chengdu 610059, China (e-mail: jpeifan@qq.com).}
\begin{abstract}
Electroencephalogram (EEG) signals are pivotal in clinical medicine, brain research, and neurological disorder studies. However, their susceptibility to contamination from physiological and environmental noise challenges the precision of brain activity analysis. Advances in deep learning have yielded superior EEG signal denoising techniques that eclipse traditional approaches. In this research, we deploy the Retentive Network architecture—initially crafted for large language models (LLMs)—for EEG denoising, exploiting its robust feature extraction and comprehensive modeling prowess. Furthermore, its inherent temporal structure alignment makes the Retentive Network particularly well-suited for the time-series nature of EEG signals, offering an additional rationale for its adoption. To conform the Retentive Network to the unidimensional characteristic of EEG signals, we introduce a signal embedding tactic that reshapes these signals into a two-dimensional embedding space conducive to network processing. This avant-garde method not only carves a novel trajectory in EEG denoising but also enhances our comprehension of brain functionality and the accuracy in diagnosing neurological ailments. Moreover, in response to the labor-intensive creation of deep learning datasets, we furnish a standardized, preprocessed dataset poised to streamline deep learning advancements in this domain. 
\end{abstract}



\maketitle
\section{Introduction}\label{}
Electroencephalography (EEG) measures neural activity as potential fluctuations on the scalp, primarily emanating from the brain's gray matter \cite{zhang2020eegdenoisenet} \cite{qin2018high}. This neural activity is detected via an array of electrodes strategically placed on the scalp \cite{turnip2014removal} \cite{tiwari2023automatic}. Analyzing EEG data yields a broad range of physiological, psychological, and pathological insights \cite{sun2020novel}. Yet, the high temporal resolution characteristic of EEG signals renders them vulnerable to diverse and complex noise sources such as cardiac, ocular, and muscular artifacts, as well as environmental interference \cite{jiang2019removal}. The prevalent intrusion of these noises during the acquisition phase significantly hampers the isolation of unadulterated EEG signals, thus severely restricting advancements in EEG-related research and practical applications \cite{molla2012artifact} \cite{tiwari2019multiclass}. Consequently, there is an imperative need for a robust EEG denoising technique that effectively reduces noise without compromising critical signal information, which is vital for advancing EEG research.

A multitude of traditional denoising techniques for EEG signal enhancement has been advanced, encompassing both regression-based and adaptive filter-based methodologies. Regression-based strategies involve estimating the noise component with a pre-established noise model and subsequently subtracting this estimate to purify the EEG data, thereby producing a cleaner signal \cite{mcmenamin2010validation} \cite{gratton1983new}. Conversely, adaptive filtering operates on a fundamentally different principle, adjusting filter coefficients in real-time based on incoming EEG data to attenuate noise \cite{he2004removal} \cite{klados2011reg}. However, these conventional techniques have their limitations. The tuning of hyperparameters in these methods critically affects the denoising performance, requiring expert judgment to optimize settings. Moreover, there is a risk of losing vital EEG information during noise reduction, which could detrimentally affect further analytical work.
 
The EEG signal is a complex waveform characterized by nonlinear features crucial for its analysis. Therefore, denoising methods must preserve these nonlinear features while eliminating noise \cite{sun2020novel}. The advancement in computer processing power and the expansion of EEG datasets \cite{zhang2020eegdenoisenet} have spurred recent research endeavors to leverage deep learning for EEG signal denoising. Commonly employed architectures for EEG denoising networks include feedforward neural networks (FNN) \cite{bebis1994feed} \cite{yang2018automatic}, convolutional neural networks (CNN) \cite{albawi2017understanding} \cite{sun2020novel}, and recurrent neural networks (RNN) \cite{zaremba2014recurrent}, along with their variations, such as long and short-term memory networks (LSTM) \cite{memory2010long} \cite{zhang2020eegdenoisenet}. As EEG is collected in the time dimension, establishing a temporal relationship between sampling points, basic network architectures have demonstrated significant improvement in denoising performance compared to traditional methods. However, they face challenges either in retaining temporal information or lacking global modeling capability while preserving temporal information. To address this, some studies have explored integrating the Transformer model \cite{vaswani2017attention} into EEG denoising tasks, as it \cite{pu2022eegdnet} effectively preserves temporal information and enables efficient data parallel computation, yielding notable results. In EEG signal processing, it is critical to accurately capture temporal order information, as patterns and features in these signals are often synchronized with specific neurophysiological events. Although the Transformer provides a powerful algorithm to process sequential data through an attentional mechanism, it does not explicitly encode the temporal order of the signals, but rather introduces this information indirectly through positional encoding. This may lead to insufficient sensitivity to the temporal dynamics of EEG signals in some cases.

In recent years, with the rapid advancement of large language model (LLM), a novel network called Retentive Network \cite{sun2023retentive} has emerged. Retentive Network gains its intuitive understanding of the temporal order in a sequence by introducing the decay mask Retention mechanism. With this mechanism, the input sequences are made naturally chronologically sequential, thus providing an effective way to model the inherent temporal dynamics of EEG signals. This approach is more suitable for processing EEG data as it is able to capture and utilize changes in bioelectrical activity over time, and thus may perform better than the standard Transformer model in practical applications. Retentive Network exhibits a favorable disposition towards temporal information, boasts robust global modeling capabilities for nonlinear features, and demonstrates commendable performance. However, when applied directly to denoise EEG signals, a challenge arises. This stems from the fact that EEG signals possess temporal characteristics and encompass global nonlinear features. Unfortunately, using Retentive Network directly for EEG denoising is unfeasible due to a misalignment in the dimensional requirements. Retentive Network, designed for two-dimensional input, conflicts with the one-dimensional nature of EEG signals. Unlike the approach in \cite{pu2022eegdnet}, reshaping a 1D signal into a 2D format results in a fixed sum of input dimensions after reshaping, compromising subsequent network feature extraction. To address this issue, we propose a signal embedding method capable of transforming the signal into a sequence of arbitrary length and embedding dimensions, enhancing network flexibility. Additionally, while EEGdenoiseNet \cite{zhang2020eegdenoisenet} introduces a standard deep learning EEG dataset, expediting the development of EEG denoising methods, the dataset remains unprocessed, lacking sample pairs. This necessitates mixing various noise types (muscle artifacts and eye artifacts) during preprocessing. Divergent data preprocessing approaches may yield disparate network results, impeding the comparison of methodologies. To mitigate this, we curated an open-source dataset using the huggingface datasets library \cite{lhoest2021datasets} from preprocessed data.
 \footnote{\href{https://github.com/woldier/EEGDiR}{https://github.com/woldier/EEGDiR}}
 \footnote{\href{https://huggingface.co/datasets/woldier/eeg\_denoise\_dataset}{https://huggingface.co/datasets/woldier/eeg\_denoise\_dataset}}
 
The main contributions of this paper can be summarized as follows:
 \begin{enumerate}[(1)]
	\item \textbf{Proposal of Signal Embedding.} In order to efficiently extract features from EEG signals, we introduce a method called signal embedding, which adds an embedding dimension to EEG signals. This method achieves the enhancement of feature information of EEG signals through the embedding strategy. The introduction of this method not only enhances the adaptability of the network, but also positively impacts the overall improvement of the system performance.

	\item \textbf{Introducing Retentive Network for EEG Signal Denoising.} For the first time, we introduce the Retentive Network architecture into the field of EEG signal denoising to address the temporal nature of EEG signals, providing a new way to explore the intersection of EEG signals and natural language processing and expanding the scope of related research. It provides a new way to explore the intersection of EEG signal and natural language processing, and also expands the scope of related research.The introduction of Retentive Network allows us to take full advantage of its time-series information friendly and global modeling, thus realizing a new denoising method for EEG signal.
	
	\item \textbf{Provide open source datasets.} When examining the standard deep learning EEG dataset provided by EEGDenoiseNet, we observe that the raw nature of the dataset and the absence of pairs of training samples hinder the comparison of different methods. To address this limitation, we create an open-source dataset using preprocessed data. This not only eliminates challenges related to noise and ensures data consistency but also facilitates the exploration of deep learning-based denoising methods for EEG signals.
\end{enumerate}









\section{Related work}\label{}
 \subsection{Traditional Methods}

Regression methods typically depend on exogenous reference channels such as EOG, EMG, or ECG \cite{yang2018automatic} to model and eliminate associated noise \cite{klados2011reg}. The efficacy of these methods is contingent upon the quality of the reference channels. Poor quality or absence of these channels significantly undermines the performance of the regression model. Furthermore, many regression methods, especially linear regression, presuppose a linear relationship between the data. However, the association between EEG signals and noise is often nonlinear, particularly when noise sources are complex, such as muscle activity or eye movement. This complexity necessitates more sophisticated nonlinear models for effective noise removal.

The Wavelet Transform method \cite{weidong2001eeg} is used to convert time-domain signals into time and frequency domains. This method is favored over the Fourier Transform due to its better tunable time-frequency tradeoff and its capability to analyze non-stationary signals. It operates by mapping the signal into the wavelet domain, where distinct properties of wavelet coefficients generated by signal and noise at various scales are utilized \cite{burger2015removal}. The primary goal is to eliminate noise-generated wavelet coefficients while preserving those from the actual signals. However, this method may lack sensitivity to the specific time-frequency characteristics of the noise in complex EEG signals.

To overcome these challenges, deep learning methods have become a promising alternative, thanks to their robust feature learning and representation capabilities, achieving significant successes in EEG denoising tasks.

 \begin{figure*}[]
	\centering
	\includegraphics[width=0.99\textwidth]{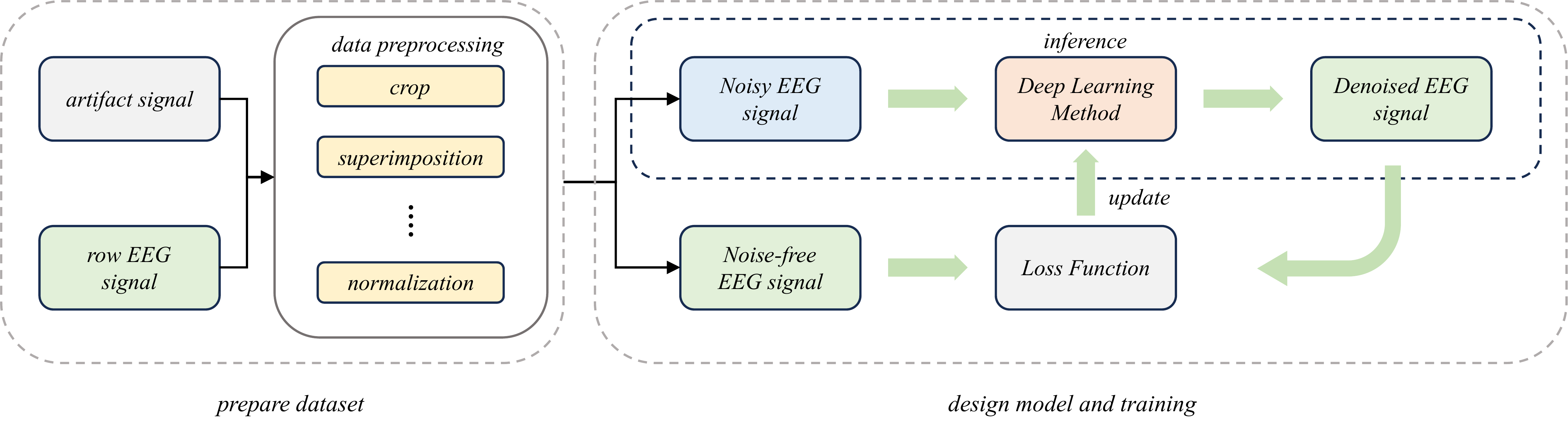}
	\caption{The diagram illustrates the training procedure of the deep learning method. It includes the Data Processing stage, where raw EEG data and various artifacts undergo preprocessing to create a suitable dataset for deep learning. The dataset comprises sample pairs, namely Noisy EEG signal and Noise-free EEG signal, representing EEG signals with and without noise, respectively. Throughout the training process, the Noisy EEG signal serves as the network input. The network, in turn, produces the Denoised EEG signal, which is utilized as the input for the subsequent training steps. The Loss Function calculates the disparity between the Denoised EEG signal and the Noise-free EEG signal, facilitating the optimization of the network. The black dashed box delineates the inference phase of the network, omitting the optimization component. This inference stage can be likened to the end-to-end output where the Noisy EEG signal is input into the network, yielding the Denoised EEG signal as the output.}\label{fig1}
\end{figure*}
  \subsection{Deep Learning Methods}

In recent years, deep learning has significantly advanced in fields such as natural language processing \cite{vaswani2017attention} \cite{sutskever2014sequence} \cite{mikolov2013efficient} and computer vision \cite{he2016deep} \cite{simonyan2014very}. Notably, its application to signal processing has demonstrated remarkable efficacy in signal denoising \cite{sun2020novel} \cite{yang2018automatic} \cite{zhang2020eegdenoisenet} \cite{pu2022eegdnet} \cite{wang2024witunet} \cite{yang2018automatic} \cite{aynali2020noise}.

To address ocular artifacts, and muscle artifacts in EEG signals, Yang et al. \cite{yang2018automatic} introduced DLN, a straightforward and efficient fully-connected neural network that surpasses traditional EEG denoising methods in processing efficiency, requiring no human intervention. Sun et al. \cite{sun2020novel} proposed a one-dimensional residual CNN (1D-ResCNN) model based on convolutional neural network CNN, showcasing superior denoising performance compared to DLN by adeptly employing various convolutional kernel sizes ($1 \times 3$, $1 \times 5$, $1 \times 7$) and integrating a residual layer \cite{he2016deep}. Zhang et al. \cite{zhang2020eegdenoisenet} presented a comprehensive EEG dataset, reducing the dataset collection challenge, and outlined four fundamental network models utilizing fully connected neural networks (FCNN), convolutional neural networks (CNN), and recurrent neural networks (RNN) for the removal of ocular and muscle artifacts. Additionally, Pu et al. \cite{pu2022eegdnet} introduced EEGDnet, leveraging the Transformer model, which outperforms prior networks in both nonlocal and local self-similarity within the model architecture. On Zhang et al.'s benchmark EEG dataset, EEGDnet surpasses previous networks in eliminating ocular artifacts, and muscle artifacts. This body of work provides valuable references and innovations to propel the advancement of EEG deep learning.

The temporal information in EEG signals is inherently long-term and characterized by numerous temporal correlations. Traditional methods often encounter difficulties in handling extensive time-series data. However, the integration of deep learning methods proves advantageous in accommodating the temporal intricacies of EEG signals. As EEG signals emanate from the entire brain, comprehensive global modeling becomes imperative for enhanced comprehension and processing. Despite the simplicity and efficiency of the DLN model, its fully-connected structure may exhibit limitations when dealing with prolonged time-series information and global modeling. While the 1D-ResCNN model surpasses DLN in denoising performance, its dependence on a single convolutional kernel size might present constraints. The model could face challenges in addressing multi-scale features and intricate temporal information. In the case of EEGDnet, its incorporation of the Transformer model demonstrates superior architectural performance concerning nonlocal and local self-similarity. However, given the diverse frequencies present in EEG signals, effective feature capture across different scales becomes crucial.
 \subsection{Dataset Preparation}

 In order to prepare data for deep learning, the raw EEG data and various artifacts undergo preprocessing, as illustrated in the left half of Fig. \ref{fig1}. This includes cropping, signal stacking, normalization, and other operations to create a dataset suitable for deep learning. The creation of this dataset is foundational to our study, as it provides immediate access to preprocessed EEG data for training deep learning models. However, data preprocessing is laborious and time-consuming, underscoring the importance of providing an out-of-the-box (i.e., no data preprocessing required) dataset readily available to researchers.
 
 During the training phase of the deep learning method, employ the framework depicted in the right half of Fig. \ref{fig1}, where noisy EEG signals and noiseless EEG signals form pairs for training. The objective of training is to model the noisy EEG signals to produce outputs closely resembling the noise-free state by learning the network's weighting parameters. Initially, the noisy EEG signal is inputted into the network to generate the corresponding denoised EEG signal. The discrepancy between this denoised EEG signal and the actual noise-free EEG signal is quantified as a loss, calculated by the loss function.
 
 The black dashed box in the figure delineates the network's inference process. During inference, the network directly takes the noisy EEG signal as input and outputs the denoised EEG signal, bypassing the optimization of the weighting parameters. This end-to-end inference capability enables the network to denoise new and unknown EEG signals in practical applications. The primary objective of the entire training process is noise suppression by optimizing the network parameters to extract true signal features amidst noise interference. This design facilitates the network to learn more effective representations, enhancing its denoising performance and providing robust support for real EEG signal processing.
  
 Given the shared structure between the training and inference processes, and provide an out-of-the-box dataset readily available to researchers. Researchers can concentrate their efforts on investigating the network's architecture. This facilitates a more profound exploration of the application of deep learning in EEG signal processing. 
 
 \section{Method}\label{}
 \subsection{Overall structure of the EEGDiR network}
   \begin{figure*}[]
 	\centering
 	\includegraphics[width=0.95\textwidth]{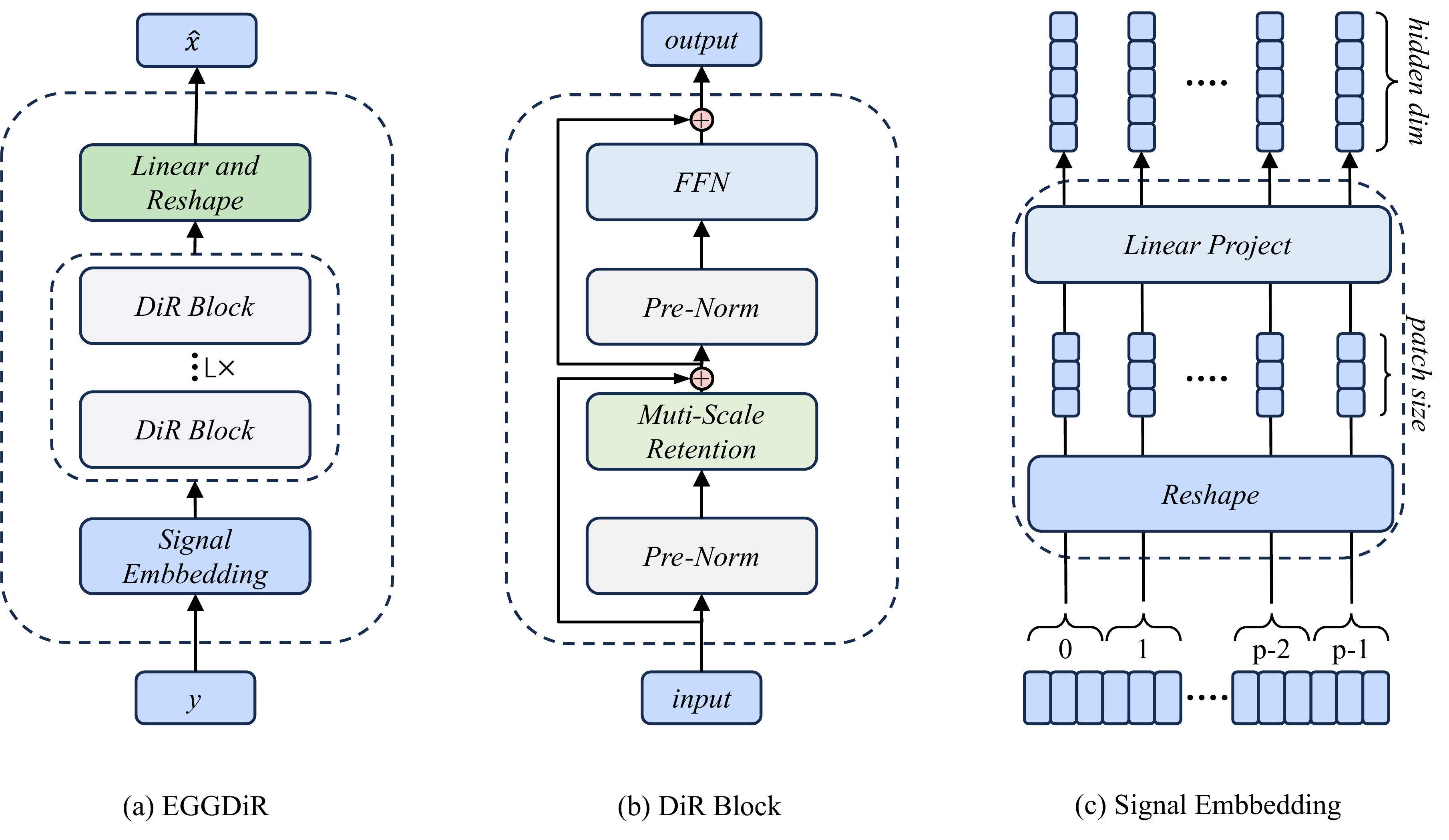}
 	\caption{(a) illustrates the architecture of the EEGDiR network. This network generates hidden dimensions through Signal Embedding and obtains the output via linear projection and transformation following multi-level DiR Block processing. EEGDiR operates as an end-to-end model, taking a noisy signal as input and producing a noise-free signal, denoted as $\hat{x}$. The DiR Block, depicted in (b), comprises Pre-Norm, Multi-scale Retention, and Residual Connection, with Pre-Norm utilizing Layer Normalization. The Signal Embedding structure, outlined in (c), involves segmenting the input sequence into new sequences based on the patch size. The hidden dimension after reshaping aligns with the patch size, and after linear projection, it matches the final hidden dimension.}
 	\label{fig2}
 \end{figure*}
 In this paper, we present EEGDiR, a novel network model tailored for EEG signal denoising. Our model incorporates Retentive Network into the realm of EEG signal denoising, introducing innovative perspectives to signal processing tasks. The overall network structure (see Fig. \ref{fig2}(a)) involves processing the noisy signal y through signal  embedding, elaborated later, to augment the embedding dimension. Subsequently, the noise signal y undergoes processing via stacked DiR Blocks at multiple levels, with the final output linearly projected to match the input dimension. EEGDiR operates as an end-to-end model, taking a noisy input signal y and generating the corresponding noiseless signal $\hat{x}$. Fig. \ref{fig2}(b) depicts the structure of the DiR Block, which begins with pre-Norm, followed by multi-scale Retention and a Residual Connection \cite{he2016deep}. By using skip connections, residual learning enables the network to learn residual mappings, which can help mitigate the degradation problem associated with increasing network depth. These residual connections provide shortcuts for gradient flow, making it easier for the network to optimize the deeper layers and improve overall performance. The output of the residual join undergoes pre-Norm once more before serving as the input for the Fully Functional Network (FFN). The term "pre-Norm" refers to Layer Normalization, employed for normalization before each submodule. Layer Normalization (LN) \cite{ba2016layer} is another normalization technique that serves as an alternative to traditional Batch Normalization (BN) \cite{ioffe2015batch}. While BN can face challenges in performance when dealing with small batch sizes, Layer Normalization aims to address these issues by normalizing at the layer level. It is crucial to note that an FFN typically includes a fully-connected module with a hidden layer that doubles the hidden dimension. The FFNs output layer reduces the hidden dimension to align with the input. The dimensions of input and output vectors remain constant across DiR and its submodules. Figure \ref{fig2}(c) illustrates the signal embedding structure, the input sequence is segmented into new sequences of length $ l_s  //  patch\_size $ based on the patch size. The initial hidden dimension of these sequences is equal to the patch size after reshaping. Through linear projection, the hidden dimension of the projected sequences matches the final hidden dimension. Given that EEG signals may encompass multiple frequencies and require effective feature capture at different scales, signal embedding is introduced to intelligently handle temporal information at varying scales. It enhances context preservation and temporal relationship retention by merging consecutive samples into a patch.
 
\subsection{Muti-Scale Retention}
 In this study, we explore the integration of the Retentive Network model, originally designed for Natural Language Processing, into the realm of EEG signal denoising. It is important to emphasize that while the Retentive Network has demonstrated remarkable success in natural language processing, our investigation centers on its applicability to EEG denoising.
The Retentive Network is comprised of L identical modules arranged in a stacked fashion, featuring residual connectivity and pre-LayerNorm akin to the Transformer architecture. Each Retnet module comprises two sub-modules: the Multi-scale Retention (MSR) module and the Feedforward Network (FFN) module. For a given input sequence $s = s_1s_2s_3\dots s_{l_s}$ (where $l_s$ denotes the length of the sequence), the input vector is initially transformed to $X^0=\left [ x_1,x_2,\dots,x_{l_x}  \right ] \in \mathbb R^{l_x\times d_{model}}$, where $d_{model}$ is the hidden dimension. Subsequently, the Retnet Block can be computed for each layer, denoted as $X^l = Retnet(X^{l-1}), l\in [1,L]$. The Simple Retention layer is defined as follows \cite{sun2023retentive}:

\begin{equation}
	\label{equation1}
	\begin{aligned}
	Q=(XW_Q)\odot \Theta,\quad K=(XW_K)\odot \bar{\Theta},\quad V=XW_V
	\end{aligned}
\end{equation}
\begin{equation}
	\label{equation2}
	\begin{aligned}
		\Theta _n =e^{in\theta },\quad D_{mn}=\begin{cases} \gamma ^{n-m},\quad n\ge m\\0,\quad n<m\end{cases}
	\end{aligned}
\end{equation}
\begin{equation}
	\label{equation3}
	\begin{aligned}
				Retention(X)=(QK^T\odot D)V
	\end{aligned}
\end{equation}
 where $\bar{\Theta}$ is the complex conjugate of $\Theta$ \cite{sun2022length} \cite{su2024roformer}, $W_Q,W_K,W_V \in  \mathbb R^{l_x\times l_x}$, $D\in\mathbb R^{l_x\times l_x}$ combines causal masking and exponential decay of relative distances into one matrix.
 
 To achieve a multichannel-like effect, input sequences can be projected to lower dimensions $d$ times, akin to the multiple-header mechanism in Transformer. This method is employed in each Retention layer with multiple headers $h = \frac{d_{model}}{d} $, where $d$ represents the length of the sequences in each header. Each header utilizes distinct  $W_Q,W_K,W_V \in  \mathbb R^{d\times d}$, constituting the Muti-Scale Retention (MSR) block. Different $\gamma $ hyperparameters are assigned to various heads in MSR, maintaining simplicity with the same $\gamma $ across different layers. Additionally, a swish \cite{jahan2023self} gate is incorporated to enhance nonlinear features in each layer. Given an input $X$, the mathematical representation of the Muti-Scale Retention is provided as follows:
\begin{equation}
	\label{equation4}
	\begin{aligned}
		 \gamma = 1 - 2^{-5-arrange(0,h)} \in R^h
	\end{aligned}
\end{equation}
\begin{equation}
	\label{equation5}
	\begin{aligned}
		head_i = Retention(X,\gamma_i)
	\end{aligned}
\end{equation}
\begin{equation}
	\label{equation6}
	\begin{aligned}
		Y = GroupNorm_n(Concat(head_1,\dots,head_h ))
	\end{aligned}
\end{equation}
\begin{equation}
	\label{equation7}
	\begin{aligned}
		MSR(X)= (swish(XW_G)\odot Y)W_o
	\end{aligned}
\end{equation}
	Here, $W_G,W_O \in\mathbb R^{d_{model}\times d_{model}} $ are learnable parameters, and GroupNorm normalizes the output of each head. Group Normalization (GN) \cite{wu2018group} is an alternative to traditional Batch Normalization (BN) \cite{ioffe2015batch} that addresses the issue of poor performance with small batch sizes. By normalizing within groups, GN provides a more robust estimation of statistics and helps mitigate the negative impact of small batch sizes or imbalanced data distribution.

\subsection{Signal Embedding}
	In our extended investigation, it was observed that when the input sequence $s=s_1s_2s_3 \dots s_{l_s}\in \mathbb R^{l_s\times 1}$ is relatively short, direct embedding is feasible. However, in general scenarios, where the time-series information of the signal is usually lengthy, direct embedding incurs high computational complexity, hindering effective network training. To address this, we propose the introduction of a concept termed "patch", involving the amalgamation of a series of consecutive samples into a single input feature. This concept is inspired by speech signal processing, where a solitary sample may inadequately represent the current word, while a segment of samples offers more semantic expressiveness. It is noteworthy that EEG signals frequently encompass extensive temporal information, and signal embedding intelligently captures this temporal data. By grouping consecutive samples into patches, the network better retains context and temporal relationships in the signal, enhancing denoising effectiveness. This approach aligns with speech signal processing, where context is pivotal for accurate speech comprehension. Consequently, this paper introduces signal embedding, a more efficient process tailored to the characteristics of EEG signals. 
	
	The complete signal embedding process is illustrated in Figure \ref{fig2}(c). Assuming a given patch size, the original sequence is divided accordingly, reducing the sequence length to $l_s//patch\_size$. Subsequently, each patch undergoes reshaping and linear projection to attain the desired hidden dimension. This process not only mitigates computational complexity but also preserves timing information more effectively. It's crucial to note that the signal embedding used here does not employ positional encoding. This is because the Retention mechanism already incorporates positional encoding considerations, obviating the need for additional positional encoding. Mathematically, it can be expressed as follows:
\begin{equation}
	\label{equation8}
	\begin{aligned}
		{s}' =Pathchfiy(s)={s}'_1{s}'_2{s}'_3\dots {s}'_{\frac{l_s}{p} } \in \mathbb{R}^{\frac{l_s}{p} \times  p} 
	\end{aligned}
\end{equation}
\begin{equation}
	\label{equation9}
	\begin{aligned}
		X^0=Embedding({s}';\omega )=[x_1,x_2,\dots,x_{|x|}] \in \mathbb{R}^{\frac{l_s}{p}\times d_{model} } 
	\end{aligned}
\end{equation}
\begin{equation}
	\label{equation10}
	\begin{aligned}
		X^0=SignalEmbedding(s;\omega )
	\end{aligned}
\end{equation}
Equation \eqref{equation8} delineates the patching process, wherein the original sequence $s$ is segmented into smaller sequences through patching, with each patch serving as an input feature. This operation not only preserves the feature information of the input (EEG signal) but also significantly truncates the length of the input sequence, thereby diminishing the computational complexity of subsequent operations. Here, ${s}'$ denotes the sequence post the patch operation. In Equation \eqref{equation9}, we illustrate the feature embedding, signifying that after the patch sequence ${s}'$, linear projection of the feature size results in the generation of the larger feature size $X^0$. This dispersion of signal features is conducive to the subsequent network's extraction of diverse features. This process allows the signal features to be spread out, facilitating the network in extracting distinct feature types. Here, $\omega$ denotes the learnable parameter, and $X^0$ remains consistent with the preceding section. If we conceptualize patching and embedding as an end-to-end operation, it can be expressed as \eqref{equation10}. In other words, the input ${s}'$ can be derived from the signal embedding module to yield $X^0$, with $\omega$ serving as the learnable parameter, akin to \eqref{equation9}.

\section{Experiments and results}\label{}
\subsection{Preliminary}
Signals disturbed by noise are acquired through the linear combination of the electrooculogram (EOG) or electromyogram (EMG) with the pristine electroencephalogram (EEG). This procedure can be mathematically represented as  Equation \eqref{equation11} \cite{zhang2020eegdenoisenet}. The mixed EEG noise signal is denoted as $y \in \mathbb{R}^{l_y}$, where $y$ represents the sequence length. The noise-free EEG signal, denoted as $x \in \mathbb{R}^{l_x}$, serves as the ground truth, and $n\in \mathbb{R}^{l_n}$ represents ocular artifacts or muscle artifacts. It is important to note that the lengths of each sequence $l_x,l_y,l_n$ are equal. To control the noise level during mixing, we introduce the hyperparameter $\lambda$, regulating the signal-to-noise ratio (SNR) of the noisy signal. Adjustment of different $\lambda$ values enables effective control of SNR magnitude to adapt to various noise environments. The SNR is calculated using Equation \eqref{equation12}, while $\lambda$ is determined by Equation \eqref{equation13}, where $RMS(\cdot)$ denotes the root mean square of the sample, $RMS(x)$ is the root mean square of the noiseless EEG signal $x$, and RMS $RMS(\lambda \cdot n)$ is the root mean square of the mixed noise $\lambda \cdot n$. These formulas provide flexible adjustment of the signal-to-noise balance to meet diverse signal quality requirements in specific application scenarios.
\begin{equation}
	\label{equation11}
	\begin{aligned}
		y=x+\lambda \cdot n
	\end{aligned}
\end{equation}
\begin{equation}
	\label{equation12}
	\begin{aligned}
		SNR = 10log(\frac{RMS(x)}{RMS(\lambda \cdot n)} )
	\end{aligned}
\end{equation}
\begin{equation}
	\label{equation13}
	\begin{aligned}
		\lambda = \frac{RMS(x)}{RMS(n)\cdot (\frac{SNR}{10})^{10}} 
	\end{aligned}
\end{equation}

In the context of deep learning applied to EEG signal denoising, the denoising process can be conceptualized as a nonlinear mapping function. This function, denoted as $\hat{x}=F(y;\theta)$, maps the EEG signal $y$ with noise to the corresponding noise-free signal $\hat{x}$. Here, $F(\cdot)$ represents the nonlinear mapping function, our neural network model, and $\theta$ is the model's learnable parameter. To facilitate better parameter learning, we employ the mean square error (MSE) as the loss function. The MSE is defined by calculating the squared difference between the predicted value $\hat{x}_i$ and the true value $x_i$ for each sample point $i$ of the signal, summing these differences, and dividing by the number of samples $n$. Mathematically, this is expressed as Equation \eqref{equation14}.
\begin{equation}
	\label{equation14}
	\begin{aligned}
			\mathbb{L}(x,\hat{x}) = \frac{1}{n}  \sum_{i=1}^{n}(x_i-\hat{x}_i )^2 
	\end{aligned}
\end{equation}

\subsection{Experiments Detail}
\subsubsection{Datasets}
To assess the denoising efficacy of the proposed EEGDiR model, we utilized the EEGDenoiseNet dataset [5], a widely adopted dataset in deep learning for EEG signal denoising, for both training and testing. The dataset encompasses various signal categories, including 4515 pristine EEG signals, 3400 ocular artifacts, and 5598 muscle artifacts. Each sample has a sampling time of 2 seconds at a rate of 256 samples per second. Pure EEG signals are denoted as $x$ in Equation \eqref{equation11}, while ocular artifacts or muscle artifacts are denoted as $n$ in Equation \eqref{equation11}.

For signals contaminated with ocular artifacts, 3400 samples were randomly chosen from pure EEG signals and all 3400 ocular artifact signals. Subsequently, the training and test sets were constructed in an 8:2 ratio, respectively. At specified signal-to-noise ratio (SNR) levels (-7 dB to 2 dB), pure EEG signals were linearly combined with ocular artifacts to generate ocular artifact-contaminated signals $y$. Notably, the parameter $\lambda$ for superimposing eye movement artifacts in Equation \eqref{equation13} was directly calculated based on the given SNR value to obtain $y$. This dataset is referred to as the EOG dataset.

The signals contaminated with muscle artifacts were derived from pure EEG signals, and all 4515 samples were utilized along with the 5598 samples from the EOG artifact signals. To maintain consistency in the number of samples from pure EEG signals and EMG artifact signals, some samples from the pure EEG signal were reused. The resulting dataset was partitioned into training and test sets in an 8:2 ratio. Similarly, using specified SNR levels, pure EEG signals were randomly combined with EMG artifacts to generate EMG artifact-contaminated signals $y$. This dataset is denoted as the EMG dataset.

\begin{figure}[!t]
	\centering
	\subfloat[]{
		\includegraphics[width=0.49\linewidth]{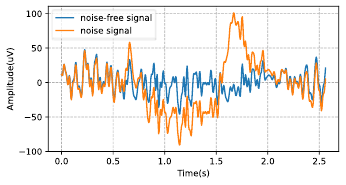}
		\label{new-3a}
	}
	\subfloat[]
	{
		\includegraphics[width=0.485\linewidth]{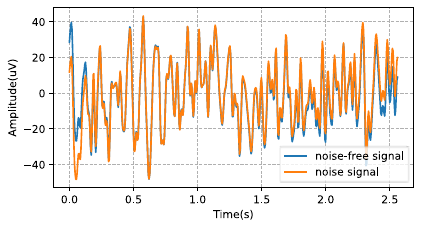}
		\label{new-3b}
	} \\
	\subfloat[]
	{
		\includegraphics[width=0.49\linewidth]{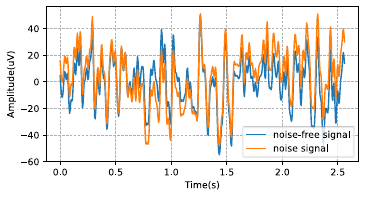}
		\label{new-3c}
	}
	\subfloat[]
	{
		\includegraphics[width=0.485\linewidth]{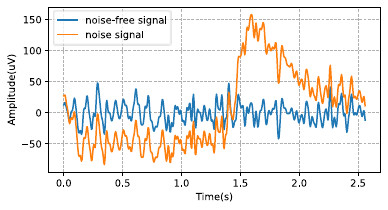}
		\label{new-3d}
	}
	\caption{The recorded signals from the first electrode of Participant 1 in SS2016 are displayed in sequence as segments 1 through 4, denoted as (a), (b), (c), and (d) respectively. Each segment comprises 512 samples, with a sampling rate of 200 SPS.}
	\label{new-3}
\end{figure}

To validate our proposed network thoroughly, we utilize a semi-simulated EEG dataset (SS2016) \cite{klados2016semi}, contaminated with ocular artifacts, alongside clean EEG signals. This dataset is notable for containing both clean EEG signals and their contaminated counterparts. Collected from $54$ participants during a closed-eye experiment, the signals are devoid of ocular artifacts. Electroencephalogram (EEG) signals were recorded from $19$ electrodes placed according to the International 10-20 system, with each experiment lasting approximately 30 seconds and sampled at 200 SPS\cite{mashhadi2020deep}.
The dataset, comprising pure and contaminated EEG signals, comprises $54$ matrices, each corresponding to one participant. Each matrix consists of $19$ channels, each representing the signal recorded by an electrode. The number of sampling points for channel signals ranged from 5600 to 8400. Data synthesis involved overlaying EOG data onto EEG signals, resulting in recordings of about $30$ seconds in duration but with varying sample counts. For ease of processing, we segmented the data without overlap into small segments of $512$. Post-segmentation, we obtained 11495 samples, with Figure \ref{new-3} displaying segments 1-4 for participant $1$.

Subsequently, we calculated the SNR values of each sample, rounding them, and depicted the results in Figure \ref{new-4}. Upon observation, we noted SNR values ranging between -10 to 20, with a relatively large number of samples falling within the -5 to 15 range, and fewer samples at other SNR values, posing challenges for network learning. To address this, we followed the data processing approach of the EEGDenoise dataset, separating the noise (n) from the pure signal (x) and contaminated signal (y). We then multiplied n by different $\lambda$ values to attain noise levels ranging from -7dB to 2dB.
However, we observed that when the signal-to-noise ratio of signal pairs in the original dataset was high, such as Figure \ref{new-3b} and Figure \ref{new-3b}, excessively large $\lambda$ values were required to achieve low signal-to-noise levels, resulting in signal amplification beyond realistic levels. Data  like this not correspond to actual noisy signals and were deemed unsuitable for training samples. Thus, we discarded pairs of samples with SNR higher than 5dB from the original dataset, resulting in a final sample count of 6716. We assigned different $\lambda$ values to n corresponding to each signal, obtaining noise levels from -7dB to 2dB. This dataset is denoted as the SS2016 EOG dataset. The resulting dataset was partitioned into training and test sets in an 8:2 ratio.

\begin{figure}[!t]
	\centering
	
	\includegraphics[width=0.7\linewidth]{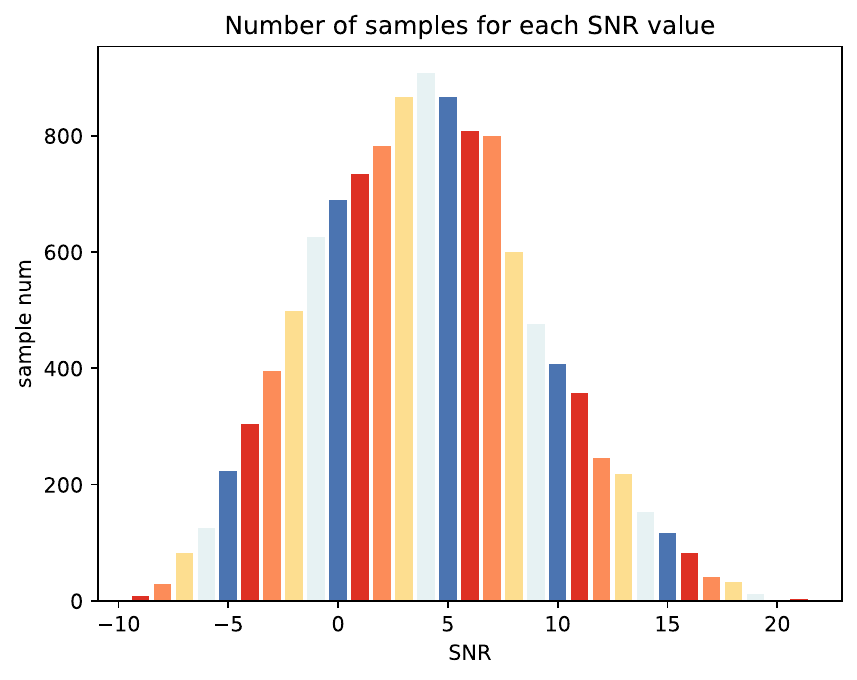}
	\label{new-4}
	
	\caption{The SNR distribution of the SS2016 dataset after signal segmentation is depicted in the figure. It's important to highlight that the SNR values are rounded up for statistical simplicity. The horizontal axis represents various SNR levels ranging from -10 to 20, while the vertical axis indicates the number of segments corresponding to each SNR level.}
	\label{new-4}
\end{figure}

In order to facilitate the learning procedure, we normalized the input contaminated EEG segment and the ground-truth EEG segment by dividing the standard deviation of contaminated EEG segment according to Equation \eqref{equation15}, where $\sigma_y$ is the standard deviation of $y$ (artifact contaminated signal).
\begin{equation}
	\label{equation15}
	\begin{aligned}
		\hat{x}=\frac{x}{\sigma _y}, \hat{y}= \frac{y}{\sigma _y}
	\end{aligned}
\end{equation}

However, it is important to note that the EEGDenoiseNet dataset solely provides raw data, and the data provided by SS2016 is not segmented, necessitating researchers to conduct their own processing. This procedure is relatively intricate, posing inconvenience for the exploration of EEG signal denoising through deep learning methodologies. To address this challenge, this paper undertakes the preprocessing of the dataset and subsequently shares the processed dataset as an open-source resource, now accessible on the Hugging Face Hub. The primary objective of this endeavor is to facilitate researchers access to and utilization of processed data, enabling them to concentrate more on the investigation of EEG deep learning denoising methods without being encumbered by the intricacies of data processing. By providing this dataset as an open-source entity, our aim is to stimulate increased research in EEG signal processing and offer a more streamlined resource for the academic community.
\subsubsection{Train details}
In this investigation, we opted to implement the EEGDiR model utilizing the PyTorch deep learning framework, renowned for its widespread usage and adaptability. To enhance the training efficiency of the network, we employed the AdamW \cite{zhuang2022understanding} optimizer, a proficient choice for managing large-scale deep learning models. The learning rate was set to $5e-4$, and the betas parameter ranged from $(0.5, 0.9)$, with meticulous adjustment to optimize the network training process.

Throughout the training phase, the network underwent 5000 epochs to ensure comprehensive learning of dataset features. Furthermore, we configured the batch size to 1000, a standard choice that balances memory utilization and training efficacy. Notably, for accelerated training, we harnessed the computational capabilities of an NVIDIA GeForce RTX 4090 Graphics Processing Unit (GPU). Leveraging the parallel computing power of GPUs significantly augmented the speed of deep learning model training, facilitating rapid experimentation and tuning for researchers.

\subsection{Results}
\subsubsection{Comparing method}
We conducted comparative experiments to assess the efficacy of our proposed EEGDiR model against established state-of-the-art deep learning EEG denoising networks, encompassing the following models:

\begin{enumerate}[(1)]
	\item Simple Convolutional Neural Networks (SCNN) \cite{zhang2020eegdenoisenet}:
	
	Network Structure: Four 1D convolutional layers with $1\times 3$ convolutional kernels, a 1-step size, and $64$ channels.
	
	Interlayer Structure: Batch Normalization and ReLU activation functions follow each convolutional layer.
	
	Output: Features linearly projected through fully connected layers to match input dimensions.

	\item One-dimensional Residual Convolutional Neural Networks (1D-ResCNN) \cite{sun2020novel}:

	Network Structure: Utilizes three distinct convolutional kernels ($1\times 3$, $1\times 5$, $1\times7$) ResBlocks for parallel feature extraction.
	
	ResBlock Structure: Each ResBlock comprises four 1D convolutional layers, with every two forming a residual block.
	
	Interlayer Structure: Activation through Batch Normalization and ReLU functions for each residual block.
	
	Output: Concatenation of the three ResBlocks' outputs, linearly projected through fully connected layers to maintain input dimensions.
	
	\item Long Short Term Memory (LSTM) Network:
	
	A Long Short-Term Memory (LSTM) network, adapted from \cite{hochreiter1997long}, is considered the benchmark for recurrent neural networks (RNNs) \cite{zaremba2014recurrent}. LSTM is capable of learning long-term dependencies, which aids in distinguishing long-term features in noise and EEG signals. Each EEG sample is sequentially fed into LSTM cells, and the output is derived from the state of each cell through a fully-connected network.
	
	\item EEG Denoise Network (EEGDnet) \cite{pu2022eegdnet}:

	Network Structure: Incorporates a Transformer infrastructure with four Transformer layers, featuring an Attention module and a FeedForward Network (FFN) module in each layer.
		 
	Module Structure: Layer Normalization applied to the input of each module.
		 
	Output: Linear projection to maintain input dimensions.

\end{enumerate}  

These benchmark networks represent diverse EEG denoising methodologies, serving as benchmarks to validate the superiority of our proposed EEGDiR model in denoising performance. These comparisons aim to offer readers a comprehensive understanding of the EEGDiR model's performance.
\subsubsection{Evaluation measures}
We assess the denoising outcomes using three methods: Relative Root Mean Squared Error in the temporal domain ($RRMSE_{t}$), $RRMSE$ in the spectral domain ($RRMSE_{s}$), and the correlation coefficient ($CC$) \cite{zhang2020eegdenoisenet}. This selection is grounded in a profound understanding of EEG signal characteristics. Firstly, considering the temporal significance of the EEG signal, we employ $RRMSE_{t}$ to quantify the relative error between the denoised and original signals in the time domain. This method exhibits sensitivity to denoising techniques preserving temporal information. Secondly, as EEG signals encapsulate rich spectral information with research often focusing on specific frequency ranges, $RRMSE_{s}$ is employed to ensure the preservation of features in the frequency domain. Lastly, acknowledging the synergistic activities between different brain regions in EEG signals, $CC$ serves as an evaluation metric. $CC$ reflects the linear relationship between the denoised and original signals, crucial for maintaining vital information about interregional correlation. The combined use of these methods facilitates a comprehensive evaluation of denoising performance across time and frequency domains, considering their adaptability to EEG signal characteristics. The mathematical expressions for $RRMSE_{t}$, $RRMSE_{s}$, and $CC$ are represented in Equation \eqref{equation16}, \eqref{equation17}, and \eqref{equation18} respectively, where $RMS(\cdot)$ denotes root mean square, $PSD(\cdot)$ denotes power spectral density, and $Cov(\cdot)$, $Var(\cdot)$ represent covariance and variance, respectively.
\begin{equation}
	\label{equation16}
	\begin{aligned}
		RRMSE_{t}=\frac{RMS(F(y)-x)}{RMS(x)} =\frac{RMS(\hat{x} -x)}{RMS(x)}
	\end{aligned}
\end{equation}
\begin{equation}
	\label{equation17}
	\begin{aligned}
		RRMSE_{s}=\frac{RMS(PSD(F(y))-PSD(x))}{RMS(PSD(x))} =\frac{Cov(\hat{x} ,x)}{Var(\hat{x})Var(x)}
	\end{aligned}
\end{equation}
\begin{equation}
	\label{equation18}
	\begin{aligned}
		CC=\frac{Cov(F(y),x)}{\sqrt{Var(F(y))Var(x)}} =  \frac{Cov(\hat{x} ,x)}{\sqrt{Var(\hat{x})Var(x)}}
	\end{aligned}
\end{equation}

\subsubsection{Ablation study}
In this study, we pioneer the application of the Retnet network to EEG signal denoising and introduce the innovative concept of signal embedding. To assess the impact of each hyperparameter on denoising performance, we conduct ablation experiments, marking the inaugural exploration of these techniques. Our initial focus is on the influence of patch size and hidden dimension, investigating their effects on network performance. Table \ref{table1} presents quantized results for the network applied to EOG, EMG and SS2016 EOG datasets under various configurations of patch size and hidden dimension hyperparameters. Notably, we observe an enhancement in denoising performance with decreasing patch size while maintaining a consistent hidden dimension. This improvement is attributed to the impact of patched sequence length on the network's feature extraction capability—smaller patch sizes result in larger mini sequence lengths, preserving more information and thereby improving denoising effectiveness. However, a cautious approach is essential as blindly reducing patch size escalates computational complexity due to increased sequence length. Thus, a delicate balance between denoising performance and computational efficiency is imperative. Furthermore, maintaining the same patch size, an increased hidden dimension corresponds to improved denoising performance, aligning with the intuitive understanding that higher dimensionality facilitates enhanced feature extraction.
\begin{table*}[]
	\caption{The effect of patch size and hidden dim on the noise reduction performance of EEGDiR. Note mini sequence length must be $l_s//patch size$, where $l_s=512$, with $8$ Heads and $4$ DiRBlock layers.}
	\label{table1}
	\begin{tabular}{|c|c|c|ccc|ccc|ccc|}
		\hline
		\multirow{2}{*}{\begin{tabular}[c]{@{}c@{}}Patch \\ size\end{tabular}} &
		\multirow{2}{*}{\begin{tabular}[c]{@{}c@{}}Mini seq.  \\ length \end{tabular}} &
		\multirow{2}{*}{\begin{tabular}[c]{@{}c@{}}Hidden \\ dim \end{tabular}} &
		\multicolumn{3}{c|}{EOG dataset} &
		\multicolumn{3}{c|}{EMG dataset} &
		\multicolumn{3}{c|}{SS2016 EOG dataset} \\ \cline{4-12} 
		&
		&
		&
		\multicolumn{1}{c|}{$RRMSE_{t}$} &
		\multicolumn{1}{c|}{$RRMSE_s$} &
		$CC$ &
		\multicolumn{1}{c|}{$RRMSE_{t}$} &
		\multicolumn{1}{c|}{$RRMSE_s$} &
		$CC$ &
		\multicolumn{1}{c|}{$RRMSE_{t}$} &
		\multicolumn{1}{c|}{$RRMSE_s$} &
		$CC$ \\ \hline
		32 &
		16 &
		512 &
		\multicolumn{1}{c|}{0.339} &
		\multicolumn{1}{c|}{0.367} &
		0.928 &
		\multicolumn{1}{c|}{0.556} &
		\multicolumn{1}{c|}{0.561} &
		0.793 &
		\multicolumn{1}{c|}{0.357} &
		\multicolumn{1}{c|}{0.392} &
		0.932 \\ \hline
		32 &
		16 &
		256 &
		\multicolumn{1}{c|}{0.353} &
		\multicolumn{1}{c|}{0.377} &
		0.911 &
		\multicolumn{1}{c|}{0.569} &
		\multicolumn{1}{c|}{0.575} &
		0.791 &
		\multicolumn{1}{c|}{0.397} &
		\multicolumn{1}{c|}{0.466} &
		0.916 \\ \hline
		32 &
		16 &
		64 &
		\multicolumn{1}{c|}{0.382} &
		\multicolumn{1}{c|}{0.371} &
		0.909 &
		\multicolumn{1}{c|}{0.572} &
		\multicolumn{1}{c|}{0.577} &
		0.790 &
		\multicolumn{1}{c|}{0.420} &
		\multicolumn{1}{c|}{0.525} &
		0.903 \\ \hline
		16 &
		32 &
		512 &
		\multicolumn{1}{c|}{0.327} &
		\multicolumn{1}{c|}{0.361} &
		0.932 &
		\multicolumn{1}{c|}{0.532} &
		\multicolumn{1}{c|}{0.501} &
		0.807 &
		\multicolumn{1}{c|}{0.315} &
		\multicolumn{1}{c|}{0.362} &
		0.948 \\ \hline
		16 &
		32 &
		256 &
		\multicolumn{1}{c|}{0.348} &
		\multicolumn{1}{c|}{0.371} &
		0.925 &
		\multicolumn{1}{c|}{0.598} &
		\multicolumn{1}{c|}{0.573} &
		0.776 &
		\multicolumn{1}{c|}{0.357} &
		\multicolumn{1}{c|}{0.395} &
		0.932 \\ \hline
		16 &
		32 &
		64 &
		\multicolumn{1}{c|}{0.374} &
		\multicolumn{1}{c|}{0.378} &
		0.912 &
		\multicolumn{1}{c|}{0.654} &
		\multicolumn{1}{c|}{0.593} &
		0.701 &
		\multicolumn{1}{c|}{0.401} &
		\multicolumn{1}{c|}{0.443} &
		0.913 \\ \hline
	\end{tabular}
\end{table*}

Subsequently, we assess the impact of varying the number of block layers L on network performance. Table \ref{table2} displays the denoising quantization results for the model applied to EOG, EMG and SS2016 EOG datasets, with fixed parameters patch size $16$, hidden dimension $512$, and heads $8$. The observations indicate a gradual enhancement in denoising performance with an increasing number of layers. This improvement is ascribed to the benefits of residual connections, whereby a higher number of layers does not lead to overfitting. The increased network depth contributes to superior feature extraction capabilities.

Following the ablation study, optimal performance is achieved when employing a patch size of $16$, hidden dimension of $512$, $8$ heads, and $4$ layers. Consequently, this configuration is chosen as the benchmark for subsequent comparisons with other networks.

\begin{table*}[]
	\caption{The effect of the layers of EEGDiR on the noise reduction performance. Note that patch size ,hidden dim and N Heads equal to $16$, $512$ and $8$ respectively.}
	\label{table2}
	\begin{tabular}{|c|ccc|ccc|ccc|}
		\hline
		\multirow{2}{*}{Pathch size} &
		\multicolumn{3}{c|}{EOG dataset} &
		\multicolumn{3}{c|}{EMG dataset} &
		\multicolumn{3}{c|}{SS2016 EOG dataset} \\ \cline{2-10} 
		&
		\multicolumn{1}{c|}{$RRMSE_{t}$} &
		\multicolumn{1}{c|}{$RRMSE_s$} &
		$CC$ &
		\multicolumn{1}{c|}{$RRMSE_{t}$} &
		\multicolumn{1}{c|}{$RRMSE_s$} &
		$CC$ &
		\multicolumn{1}{c|}{$RRMSE_{t}$} &
		\multicolumn{1}{c|}{$RRMSE_s$} &
		$CC$ \\ \hline
		4 &
		\multicolumn{1}{c|}{0.327} &
		\multicolumn{1}{c|}{0.361} &
		0.932 &
		\multicolumn{1}{c|}{0.532} &
		\multicolumn{1}{c|}{0.501} &
		0.807 &
		\multicolumn{1}{c|}{0.315} &
		\multicolumn{1}{c|}{0.362} &
		0.948 \\ \hline
		3 &
		\multicolumn{1}{c|}{0.356} &
		\multicolumn{1}{c|}{0.380} &
		0.925 &
		\multicolumn{1}{c|}{0.578} &
		\multicolumn{1}{c|}{0.583} &
		0.789 &
		\multicolumn{1}{c|}{0.381} &
		\multicolumn{1}{c|}{0.447} &
		0.922 \\ \hline
		2 &
		\multicolumn{1}{c|}{0.372} &
		\multicolumn{1}{c|}{0.383} &
		0.917 &
		\multicolumn{1}{c|}{0.591} &
		\multicolumn{1}{c|}{0.596} &
		0.781 &
		\multicolumn{1}{c|}{0.396} &
		\multicolumn{1}{c|}{0.461} &
		0.917 \\ \hline
		1 &
		\multicolumn{1}{c|}{0.394} &
		\multicolumn{1}{c|}{0.429} &
		0.908 &
		\multicolumn{1}{c|}{0.613} &
		\multicolumn{1}{c|}{0.594} &
		0.766 &
		\multicolumn{1}{c|}{0.411} &
		\multicolumn{1}{c|}{0.478} &
		0.911 \\ \hline
	\end{tabular}
\end{table*}

\subsubsection{Denoising effect of each method at all noise levels}
Table \ref{table3} illustrates the denoising efficacy of various methods on EOG, EMG and SS2016 EOG datasets. The outcomes in this table lead to the following:
\begin{enumerate}[(1)]
	\item Due to its relatively simplistic structure comprising only four convolutional layers and lacking residual connections, SCNN exhibits suboptimal denoising effects, potentially prone to overfitting.
	\item Featuring a more intricate architecture incorporating diverse convolutional kernels for multi-scale feature extraction and alleviating overfitting through the introduction of residual connections, 1D-ResCNN surpasses SCNN, significantly enhancing denoising outcomes. Thanks to the temporal information added to the input by the LSTM structure, LSTM exhibits better denoising results than 1D-ResCNN.
	\item Leveraging the transformer architecture, EEGDnet excels in denoising, benefitting from the global modeling prowess of the attention mechanism, complemented by residual connections and layer normalization. This results in substantial denoising improvements compared to SCNN, 1D-ResCNN and LSTM.
	\item Capitalizing on the  Retentive Network, EEGDiR achieves superior denoising performance by comprehensively understanding input temporal information and exhibiting robust global modeling capabilities. The incorporation of residuals and multiple normalizations (layer norm, group norm) further distinguishes EEGDiR, outperforming other networks. Moreover, guided by our proposed signal embedding, EEGDiR intelligently processes temporal information. This strategy adeptly captures the contextual and temporal relationships within EEG signals, aligning with their prolonged temporal characteristics. The signal embedding strategy contributes to optimized denoising performance, reinforcing EEGDiR's exceptional superiority over alternative networks.
\end{enumerate}

\begin{table*}[]

	\caption{Average performances of all SNRs (from $-7$ dB to $2$ dB). The smaller $RRMSE_{t}$ and  $RRMSE_{s}$ , and the larger $CC$, the better denoising effect. Note that all the models are trained and tested on the same data set. The baseline of EEGDiR consists of $4$ layers and $8$ Heads with patch size $16$ and hidden dim $512$. For $RRMSE_{t}$ , $RRMSE_{s}$ , the lower the better. For $CC$, the higher the better. The best result is shown in bold.}
	\label{table3}
	\begin{tabular}{|c|ccc|ccc|ccc|}
		\hline
		\multirow{2}{*}{Model} &
		\multicolumn{3}{c|}{EOG dataset} &
		\multicolumn{3}{c|}{EMG dataset} &
		\multicolumn{3}{c|}{SS2016 EOG dataset} \\ \cline{2-10} 
		&
		\multicolumn{1}{l|}{$RRMSE_{t}$} &
		\multicolumn{1}{l|}{$RRMSE_{s}$} &
		\multicolumn{1}{l|}{$CC$} &
		\multicolumn{1}{l|}{$RRMSE_{t}$} &
		\multicolumn{1}{l|}{$RRMSE_s$} &
		\multicolumn{1}{l|}{$CC$} &
		\multicolumn{1}{c|}{$RRMSE_{t}$} &
		\multicolumn{1}{c|}{$RRMSE_s$} &
		$CC$ \\ \hline
		SCNN &
		\multicolumn{1}{c|}{0.6176} &
		\multicolumn{1}{c|}{0.5905} &
		0.7938 &
		\multicolumn{1}{c|}{0.7342} &
		\multicolumn{1}{c|}{0.7977} &
		0.7364 &
		\multicolumn{1}{c|}{0.5893} &
		\multicolumn{1}{c|}{0.6724} &
		0.8156 \\ \hline
		1D-ResCNN &
		\multicolumn{1}{c|}{0.5409} &
		\multicolumn{1}{c|}{0.5900} &
		0.8503 &
		\multicolumn{1}{c|}{0.6921} &
		\multicolumn{1}{c|}{0.6848} &
		0.7434 &
		\multicolumn{1}{c|}{0.5523} &
		\multicolumn{1}{c|}{0.5804} &
		0.8552 \\ \hline

		LSTM &
		\multicolumn{1}{c|}{0.5290} &
		\multicolumn{1}{c|}{0.4894} &
		0.8449 &
		\multicolumn{1}{c|}{0.6560} &
		\multicolumn{1}{c|}{0.6092} &
		0.7461 &
		\multicolumn{1}{c|}{0.4823} &
		\multicolumn{1}{c|}{0.5573} &
		0.8747 \\ \hline

		EEGDnet &
		\multicolumn{1}{c|}{0.4819} &
		\multicolumn{1}{c|}{0.4647} &
		0.8725 &
		\multicolumn{1}{c|}{0.6200} &
		\multicolumn{1}{c|}{0.5565} &
		0.7711 &
		\multicolumn{1}{c|}{0.4594} &
		\multicolumn{1}{c|}{0.5267} &
		0.8875 \\ \hline
		EEGDiR(ours) &
		\multicolumn{1}{c|}{\textbf{0.3279}} &
		\multicolumn{1}{c|}{\textbf{0.3616}} &
		\textbf{0.9329} &
		\multicolumn{1}{c|}{\textbf{0.5322}} &
		\multicolumn{1}{c|}{\textbf{0.5004}} &
		\textbf{0.8072} &
		\multicolumn{1}{c|}{\textbf{0.3146}} &
		\multicolumn{1}{c|}{\textbf{0.3613}} &
		\textbf{0.9488} \\ \hline
	\end{tabular}
\end{table*}

\subsubsection{Denoising effect of each method at different noise levels}
\begin{figure}[!t]
	\centering
	\subfloat[$RRMSE_{temporal}$]{
		\includegraphics[width=0.33\linewidth]{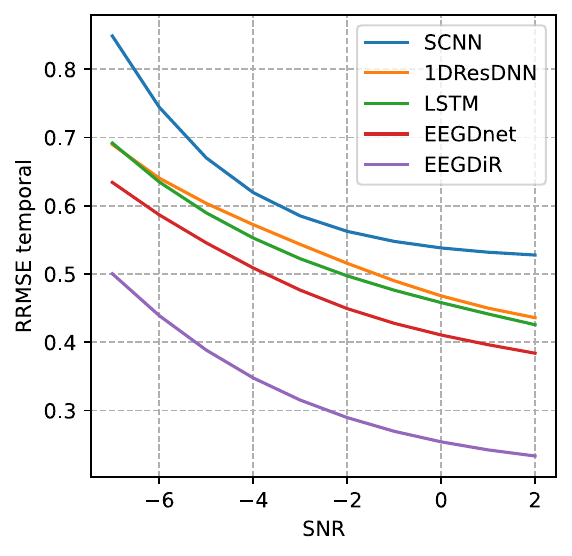}
		\label{fig3a}
	}
	\subfloat[$RRMSE_{spectral}$]
	{
		\includegraphics[width=0.33\linewidth]{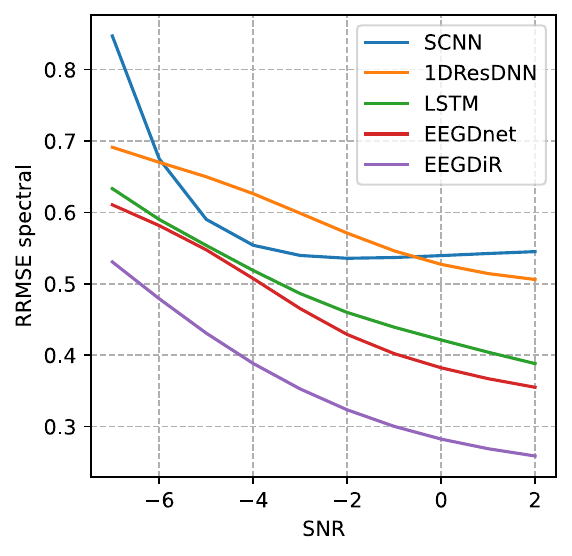}
		\label{fig3b}
	}
	\subfloat[$CC$]
	{
		\includegraphics[width=0.33\linewidth]{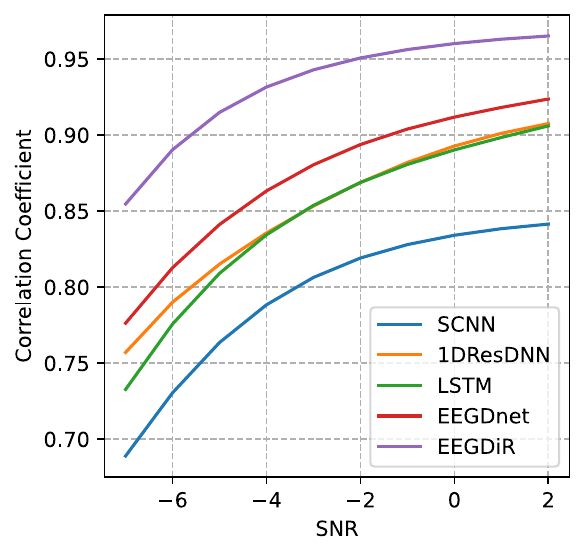}
		\label{fig3c}
	}
	\caption{Performance of four deep-learning networks at different SNR levels with EOG dataset artifact removal. The smaller $RRMSE_{t}$ and $RRMSE_{s}$ , and the larger Correlation Coefficient($CC$), the better denoising effect. The denoising performance increases as the SNR increases.}
	\label{fig3}
\end{figure}

\begin{figure}[!t]
	\centering
	\subfloat[$RRMSE_{temporal}$]{
		\includegraphics[width=0.33\linewidth]{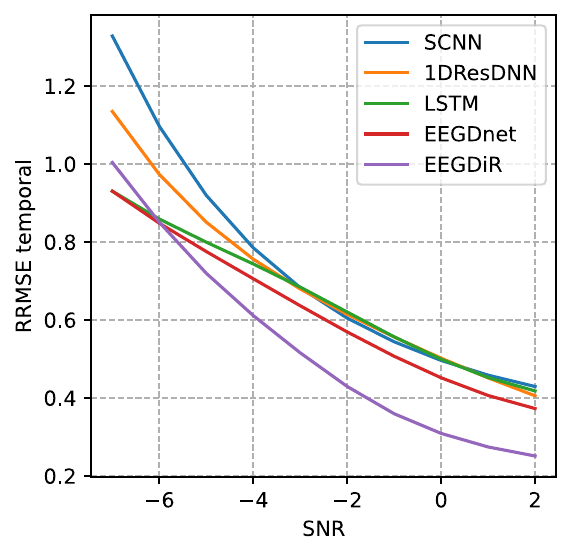}
		\label{fig4a}
	}
	\subfloat[$RRMSE_{spectral}$]
	{
		\includegraphics[width=0.33\linewidth]{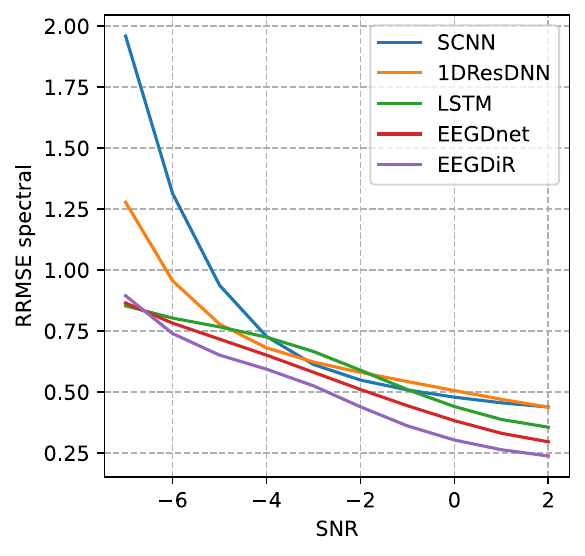}
		\label{fig4b}
	}
	\subfloat[$CC$]
	{
		\includegraphics[width=0.33\linewidth]{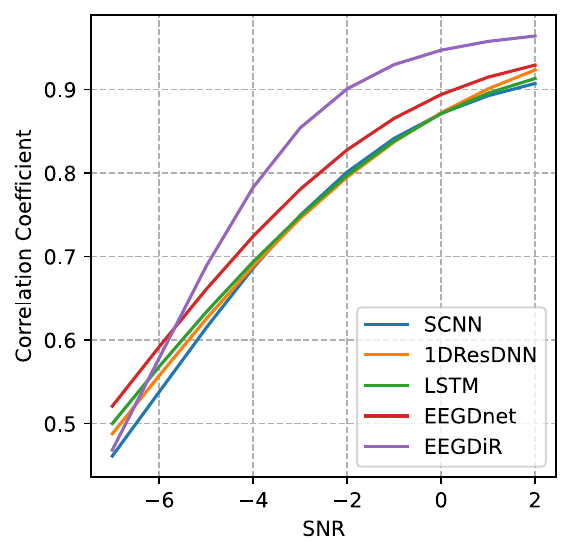}
		\label{fig4c}
	}
	\caption{Performance of four deep-learning networks at different SNR levels with EMG dataset artifact removal. The smaller $RRMSE_{t}$ and $RRMSE_{s}$, and the larger Correlation Coefficient($CC$), the better denoising effect. The denoising performance increases as the SNR increases.}
	\label{fig4}
\end{figure}

\begin{figure}[!t]
	\centering
	\subfloat[$RRMSE_{temporal}$]{
		\includegraphics[width=0.33\linewidth]{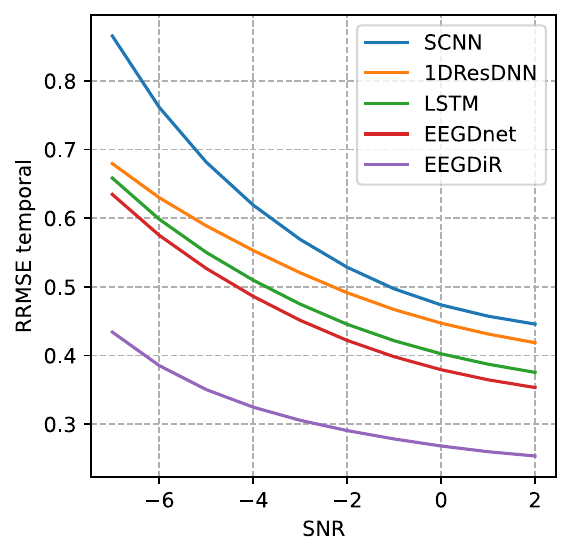}
		\label{new-1a}
	}
	\subfloat[$RRMSE_{spectral}$]
	{
		\includegraphics[width=0.33\linewidth]{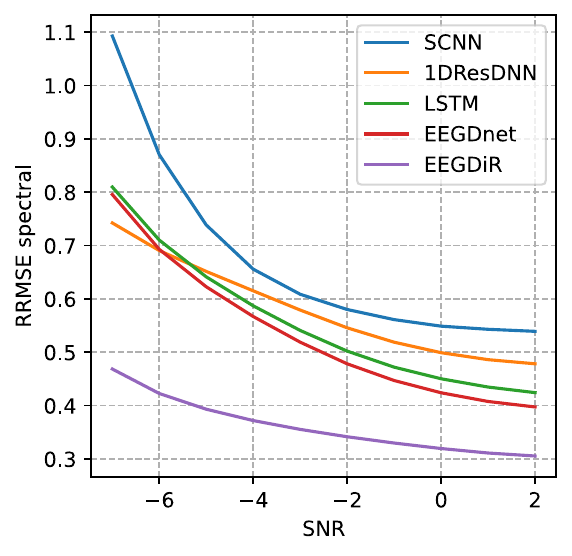}
		\label{new-1b}
	}
	\subfloat[$CC$]
	{
		\includegraphics[width=0.33\linewidth]{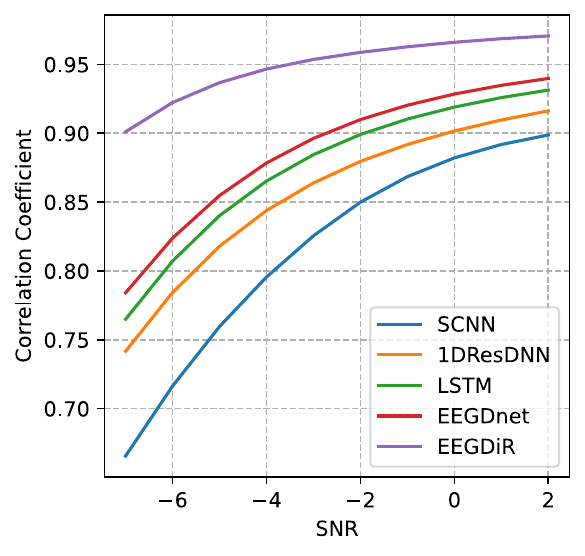}
		\label{new-1c}
	}
	\caption{Performance of four deep-learning networks at different SNR levels with SS2016 EOG dataset artifact removal. The smaller $RRMSE_{t}$ and $RRMSE_{s}$, and the larger Correlation Coefficient($CC$), the better denoising effect. The denoising performance increases as the SNR increases.}
	\label{new-1}
\end{figure}

In the subsequent section, we present the quantitative benchmarking results ($RRMSE_{t}$, $RRMSE_{s}$, $CC$ ) of diverse methods across varying SNR levels in the test set. Figures \ref{fig3}, Figures \ref{fig4} and Figures \ref{new-1} showcase the test outcomes on the EOG, EMG and SS2016 EOG test dataset, pivotal for evaluating the denoising efficacy of the methods.
\begin{enumerate}[(1)]
	\item Primarily, the performance of all methods exhibits a decline as the SNR level decreases. This negative correlation arises due to the gradual increase in noise level, posing a greater challenge for the methods in noise removal.
	\item Among the methods, SCNN displays the highest $RRMSE_{t}$ and $RRMSE_{s}$, along with the lowest $CC$. This indicates SCNN inferior denoising performance, attributed to its relatively simple network structure hindering effective input feature extraction. In contrast, the more intricate 1D-ResCNN and LSTM yields significantly improved denoising outcomes. However, compared to EEGDnet with a Transformer model and global modeling capability, there are discernible performance gaps. The EEGDiR, incorporating Rententive Network and signal embedding, achieves the lowest $RRMSE_{t}$ and  $RRMSE_{s}$, coupled with the highest $CC$. It excels in denoising tasks across varying noise levels.
	\item Analyzing the $RRMSE_{t}$ results on the EOG and SS2016 EOG dataset, denoising performance improves with decreasing noise levels and increasing SNR levels across all methods. However, the performance gap between methods persists, potentially due to EOG noise being more easily removed than EMG noise. On the EMG dataset, the performance gap diminishes as noise levels decrease (SNR levels increase), particularly evident for SCNN, 1D-ResCNN, LSTM, and EEGDnet. Nevertheless, EEGDiR maintains superior denoising performance.
	\item Evaluation of the $RRMSE_{s}$ results on the EOG and SS2016 EOG dataset indicates weaker denoising performance for SCNN, 1D-ResCNN and LSTM, possibly due to limited global modeling capability. Conversely, EEGDnet and EEGDiR exhibit superior denoising performance owing to their robust global modeling ability. On the EMG dataset, despite decreasing differences in performance as noise levels decrease, EEGDnet and EEGDiR consistently outperform SCNN, 1D-ResCNN and LSTM.
	\item Examination of $CC$ results on the EOG and SS2016 EOG dataset reveals improved denoising performance for all methods as noise levels decrease, with relatively stable performance differences. In the EMG dataset, SCNN exhibits poorer performance due to the dataset's more complex noise. Conversely, the denoising performance of the remaining three networks improves as noise levels decrease, with consistent performance differences. Notably, EEGDiR maintains excellent denoising performance throughout.
\end{enumerate}

\begin{figure}[!t]
	\centering
	\subfloat[$RRMSE_{temporal}$]{
		\includegraphics[width=0.31\linewidth]{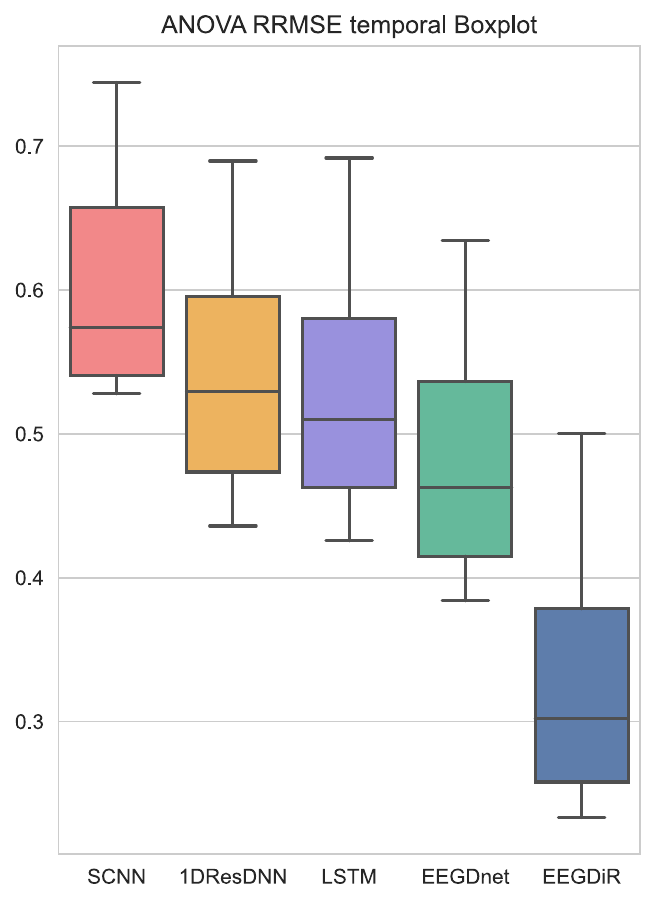}
		\label{fig5a}
	}
	\subfloat[$RRMSE_{spectral}$]
	{
		\includegraphics[width=0.31\linewidth]{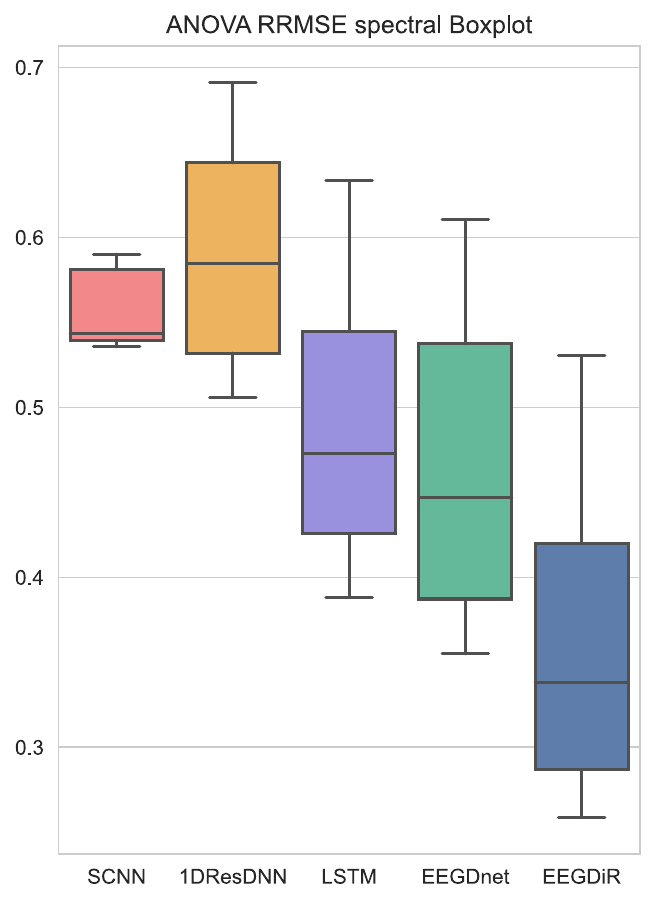}
		\label{fig5b}
	}
	\subfloat[$CC$]
	{
		\includegraphics[width=0.32\linewidth]{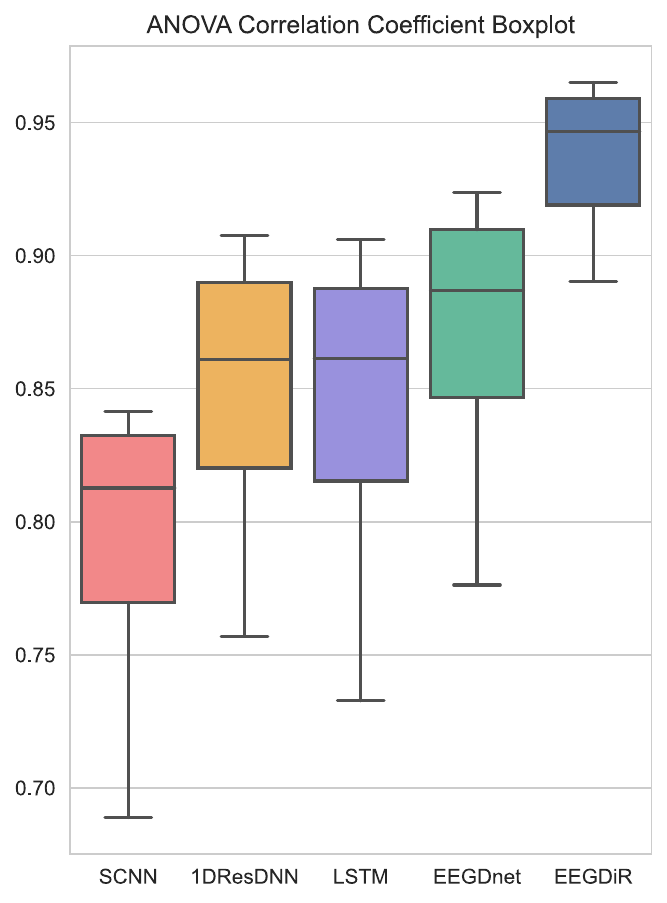}
		\label{fig5c}
	}
	\caption{Performance of four DL networks (SCNN, 1D-ResDNN, LSTM, EEGDnet, EEGDiR) in EOG dataset artifact removal. The smaller $RRMSE_{t}$ and $RRMSE_{s}$, and the larger Correlation Coefficient($CC$), the better denoising effect. EEGDiR models robustly outperform other model for EEG denosing.}
	\label{fig5}
\end{figure}

\begin{figure}[!t]
	\centering
	\subfloat[$RRMSE_{temporal}$]{
		\includegraphics[width=0.31\linewidth]{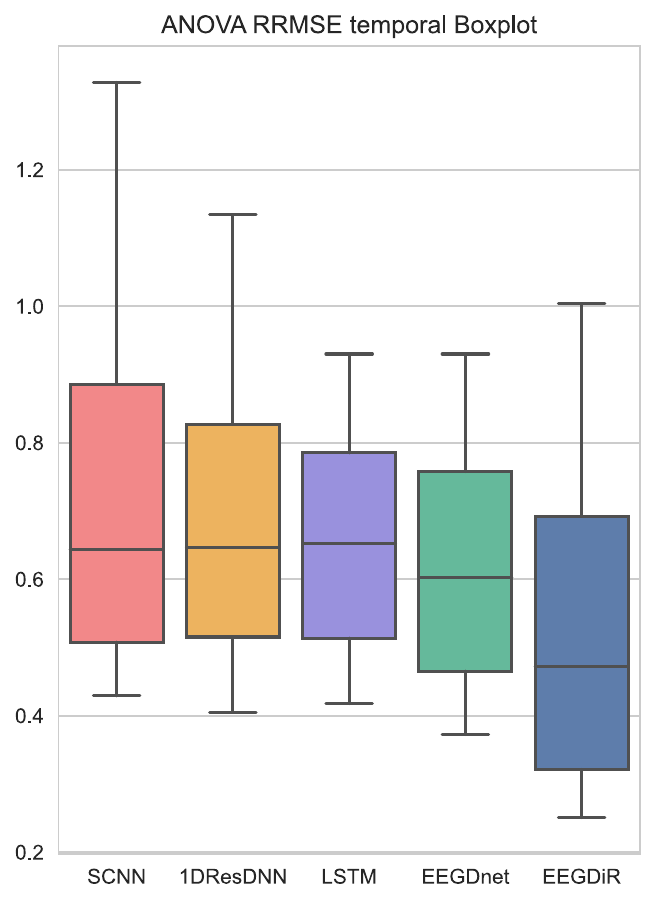}
		\label{fig6a}
	}
	\subfloat[$RRMSE_{spectral}$]
	{
		\includegraphics[width=0.315\linewidth]{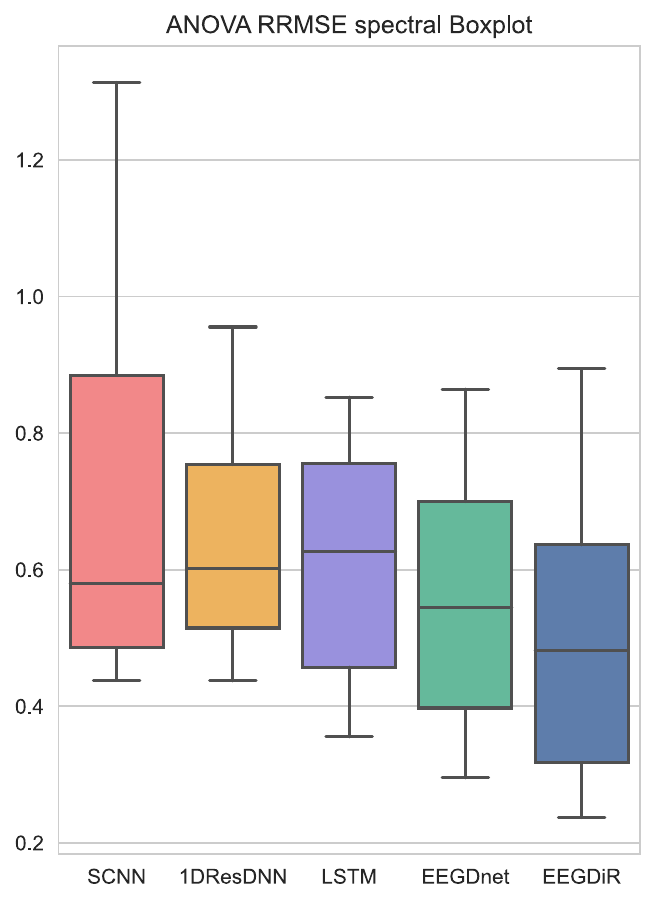}
		\label{fig6b}
	}
	\subfloat[$CC$]
	{
		\includegraphics[width=0.31\linewidth]{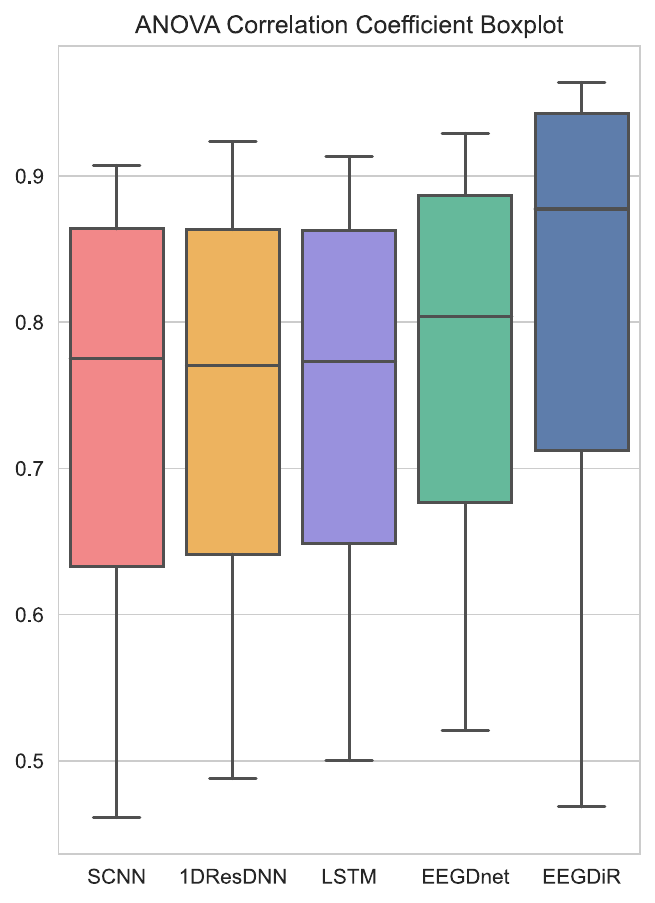}
		\label{fig6c}
	}
	\caption{Performance of four DL networks (SCNN, 1D-ResDNN, LSTM, EEGDnet, EEGDiR) in EMG dataset artifact removal. The smaller $RRMSE_{t}$ and $RRMSE_{s}$, and the larger Correlation Coefficient($CC$), the better denoising effect. EEGDiR models robustly outperform other model for EEG denosing.}
	\label{fig6}
\end{figure}

\begin{figure}[!t]
	\centering
	\subfloat[$RRMSE_{temporal}$]{
		\includegraphics[width=0.31\linewidth]{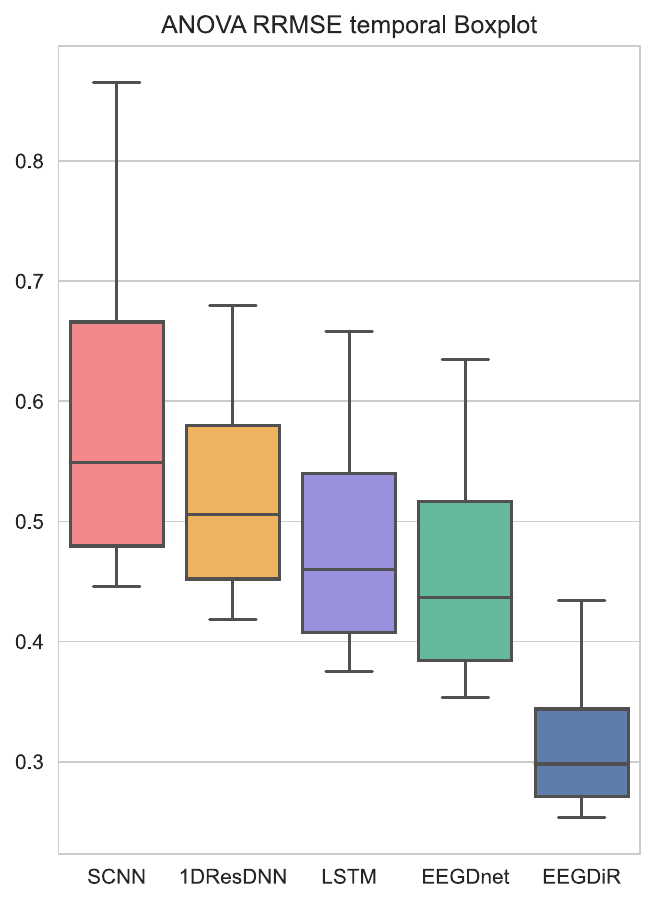}
		\label{new-2a}
	}
	\subfloat[$RRMSE_{spectral}$]
	{
		\includegraphics[width=0.31\linewidth]{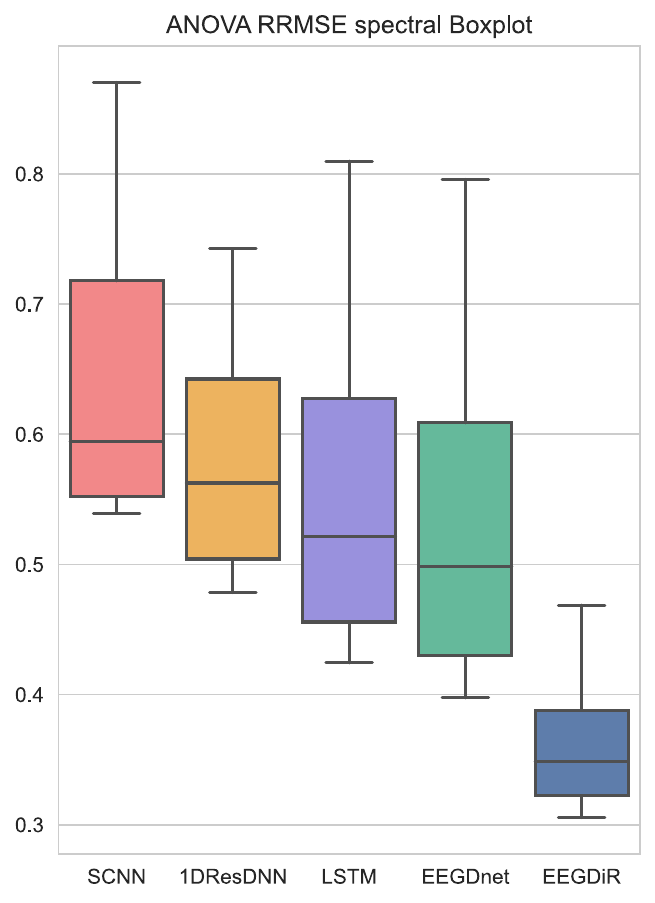}
		\label{new-2b}
	}
	\subfloat[$CC$]
	{
		\includegraphics[width=0.32\linewidth]{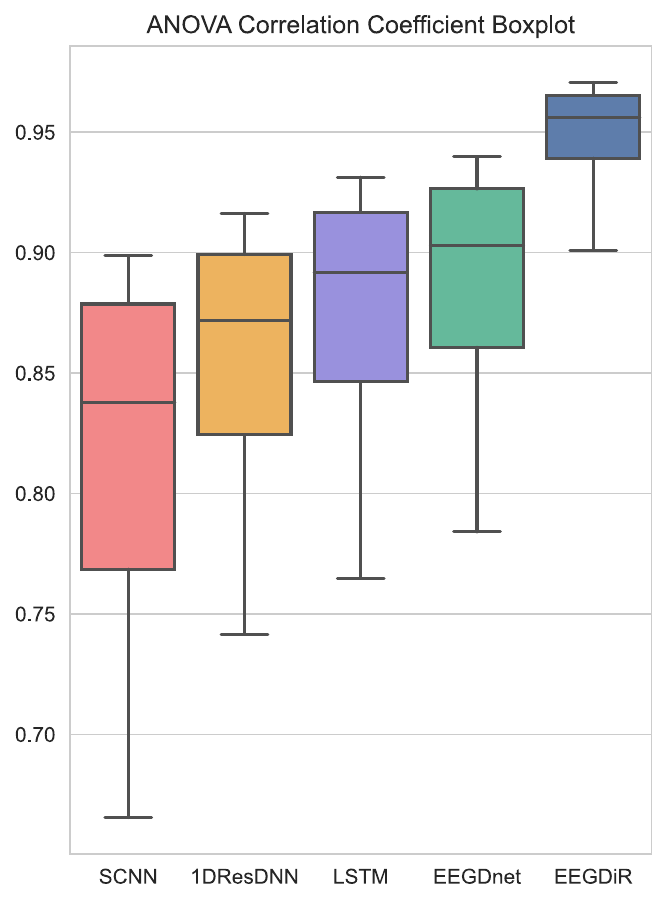}
		\label{new-2c}
	}
	\caption{
		Performance of four DL networks (SCNN, 1D-ResDNN, LSTM, EEGDnet, EEGDiR) in SS2016 EOG dataset artifact removal. The smaller $RRMSE_{temporal}$ and $RRMSE_{spectral}$, and the larger Correlation Coefficient($CC$), the better denoising effect. EEGDiR models robustly outperform other model for EEG denosing.
	}
	\label{new-2}
\end{figure}

Figures \ref{fig5}, Figures \ref{fig6} and Figures \ref{new-2} illustrate the ANOVA results for models evaluated on the EOG, EMG and SS2016 EOG datasets. Drawing conclusions from the provided information and ANOVA analyses, the following observations emerge:
\begin{enumerate}[(1)]
	\item   $RRMSE_{t}$ Rancking: The denoising performance across the four methods is observed as follows: SCNN < 1D-ResCNN < LSTM < EEGDnet < EEGDiR. ANOVA analysis indicates significant differences in $RRMSE_{t}$, with marked distinctions in the EOG dataset and relatively modest differences in the EMG dataset. EEGDiR significantly outperforms other methods in time-domain denoising for both EOG, EMG and SS2016 EOG datasets, followed by EEGDnet, LSTM and 1D-ResCNN, while SCNN exhibits the least efficacy.
	\item $RRMSE_{s}$ Rancking: The denoising performance order for $RRMSE_{s}$ is 1D-ResCNN < SCNN < LSTM < EEGDnet < EEGDiR. The occurrence of 1D-ResCNN < SCNN is attributed to 1D-ResCNN superior extraction of features in the time domain, leading to diminished denoising performance in the spectral features. ANOVA results show a significant difference in $RRMSE_{s}$ among methods on the EOG and SS2016 EOG dataset, while the difference is relatively weak on the EMG dataset. EEGDiR significantly outperforms other methods in spectral denoising, followed by EEGDne, LSTM and SCNN, while 1D-ResCNN is less effective.
	\item $CC$ Metric Ranking: The denoising performance sequence on the $CC$ metric remains SCNN < 1D-ResCNN < LSTM < EEGDnet < EEGDiR. ANOVA analysis reveals a significant difference in $CC$ between methods for the EOG and SS2016 EOG dataset, while the difference is relatively weak for the EMG dataset. Comparing mean values, EEGDiR excels in correlation, followed by EEGDnet and 1D-ResCNN, while SCNN exhibits poor CC performance.
\end{enumerate}

In summary, the outstanding denoising performance of the EEGDiR method can be attributed to multiple factors.
The Rentetive Network architecture provides enhanced global modeling capability, enabling a more accurate restoration of input timing information. The proposed signal embedding method adeptly handles the prolonged temporal information of EEG signals, capturing context and temporal relationships intelligently through the combination of successive sampling points into patches. This advantage enables EEGDiR to achieve superior denoising effects in both the time domain and spectral characteristics. Additionally, the synergy of residual connectivity and multiple normalization methods (layer norm, group norm) enhances EEGDiR denoising performance and robustness to noise. The advanced Retnet architecture, skillful embedding strategy, and enhanced network design collectively contribute to EEGDiR exceptional performance in time-domain and spectral denoising, as well as correlation.

\subsubsection{Visualization of denoising results for each method on EOG and EMG datasets}
\begin{figure}[!t]
	\centering
	\subfloat[Denoising outcomes on EOG dataset.]{
		\includegraphics[width=1\linewidth]{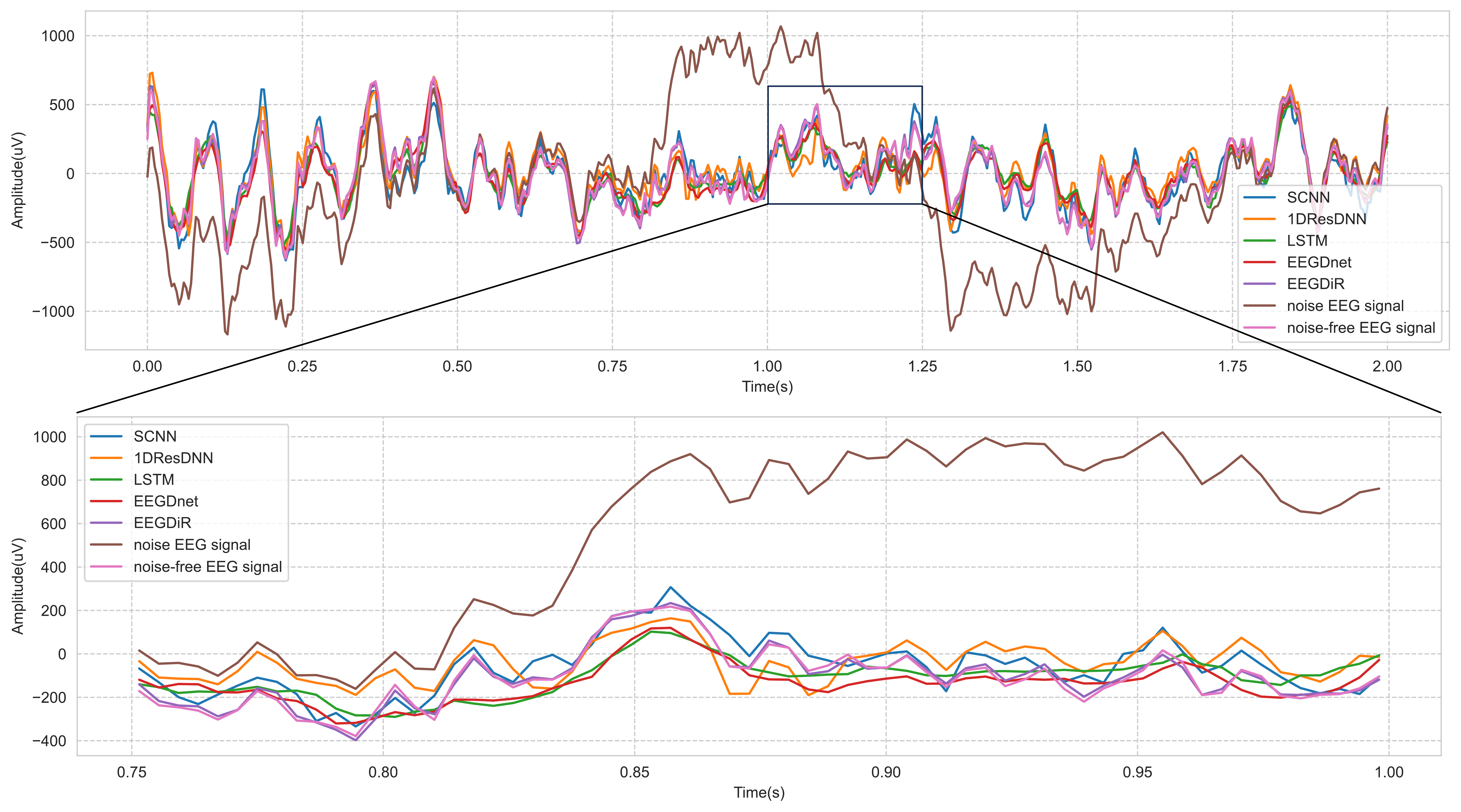}
		\label{fig7a}
	}
	\hfil
	\subfloat[Denoising outcomes on  EMG dataset.]
	{
		\includegraphics[width=1\linewidth]{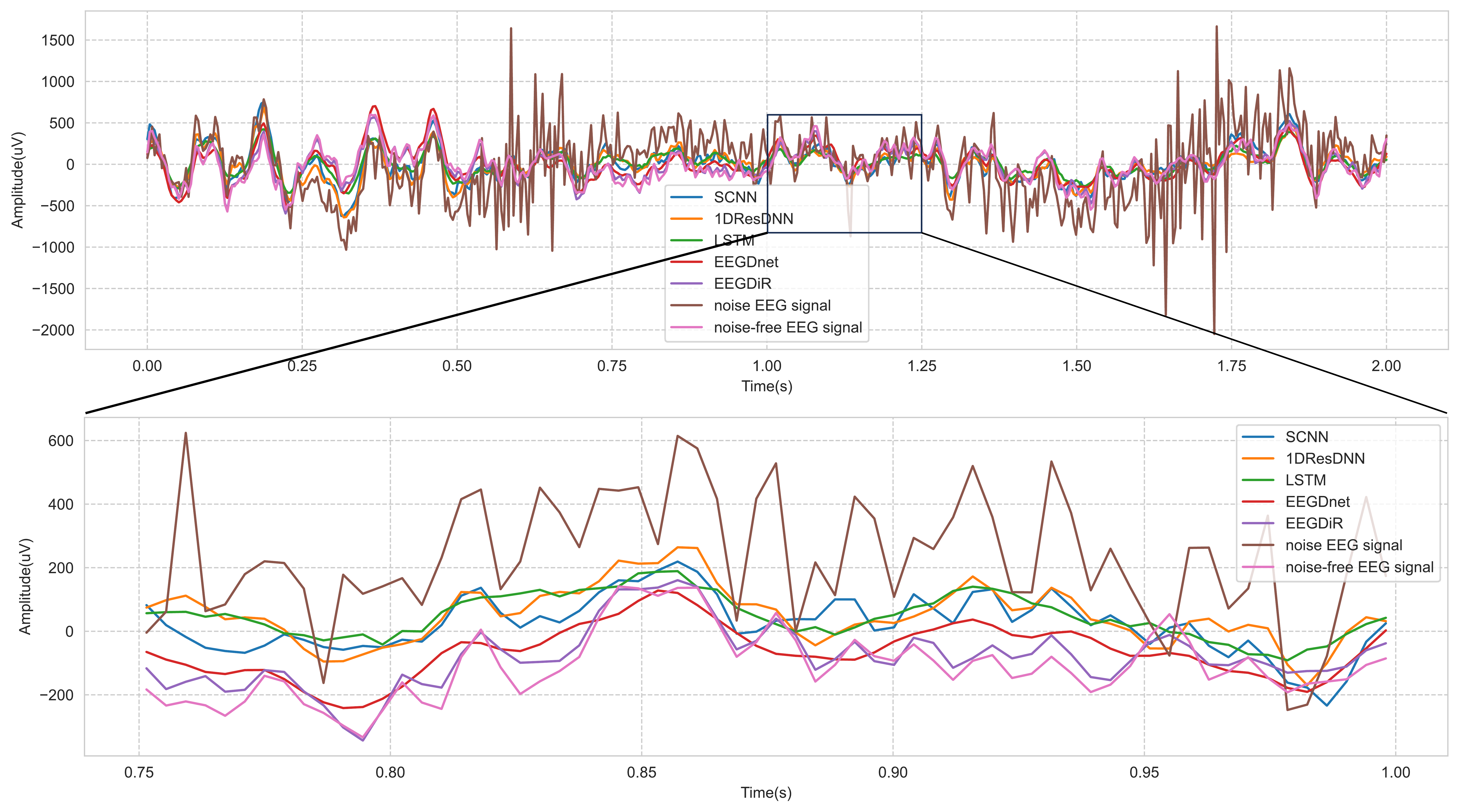}
		\label{fig7b}
	}\hfil
	\subfloat[Denoising outcomes on SS2016 EOG dataset]
	{
		\includegraphics[width=1\linewidth]{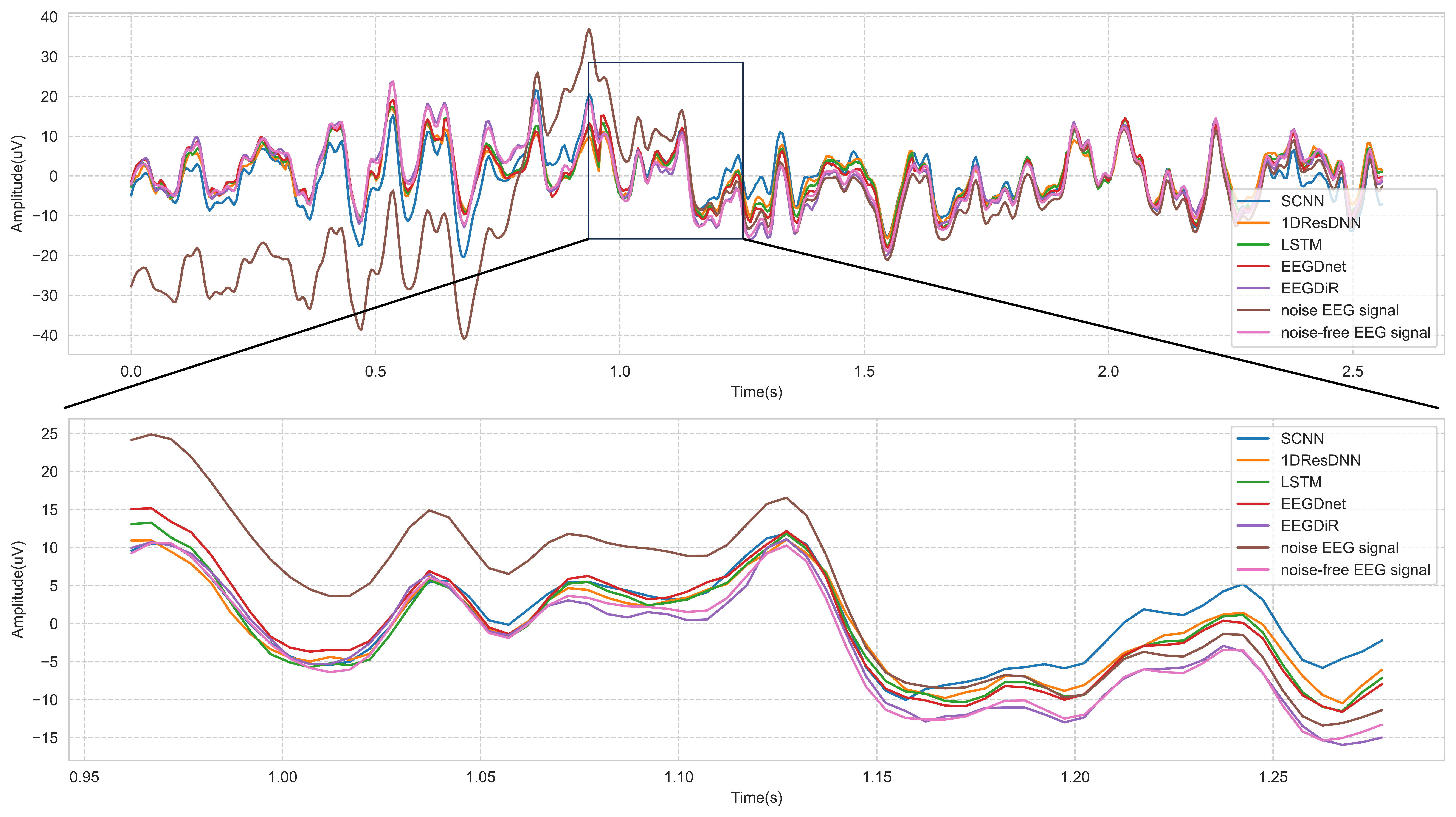}
		\label{fig7c}
	}
	\caption{Visualization of denoising outcomes for various state-of-the-art models: (a) Denoising outcomes on EOG dataset. (b) Denoising outcomes on EMG dataset. (b) Denoising outcomes on SS2016 EOG dataset. A closer inspection can be facilitated by zooming in for a more detailed view. It is important to highlight that we have restored the network output's amplitude back to the original data scale through back-normalization. The temporal domain operates at a sampling rate of 256 SPS for EOG and EMG dataset. The temporal domain operates at a sampling rate of 200 SPS for SS2016 EOG dataset. From the provided results, it is evident that the denoising results achieved by the proposed EEGDiR model in this study closely approximate the true signal.}
	\label{fig7}
\end{figure}

The visualization results depicting the impact of EOG and EMG noise on EEG signals are presented in Figure \ref{fig7a},  Figure \ref{fig7b} and Figure \ref{fig7c}, yielding the following observations. It is noteworthy that the dataset has undergone variance normalization. When presenting the visualization results, Equation \eqref{equation15} is employed to scale down the results to the original data scale, enhancing the accuracy of showcasing the noise effect on EEG signals.

\begin{enumerate}[(1)]
	\item All methods exhibit some degree of noise suppression in noisy signals, underscoring the variability among different denoising approaches. Particularly notable is the substantial difference between the denoising results of the SCNN method and the noisy signal. This divergence may be attributed to the relatively simplistic network structure of SCNN, hindering comprehensive feature extraction.
	\item The relatively complex structure and residual connectivity of LSTM and 1D-ResCNN result in an improvement in denoising compared to SCNN, emphasizing the impact of network architecture on denoising performance. Leveraging the global modeling and feature extraction capabilities facilitated by the Transformer's attention mechanism, EEGDnet outperforms SCNN, LSTM and 1D-ResCNN in denoising. This underscores the enhancement of network performance with the introduction of the attention mechanism.
	\item Overall, the denoising effect achieved by EEGDiR closely approaches that of a noise-free signal. This notable advantage can be attributed to the synergistic effect of the Retentive Network architecture and our proposed signal embedding, tailored to match the characteristics of EEG signals. Firstly, the Retentive Network architecture enhances understanding of timing information in EEG signals through its robust global modeling capability. This enables the network to accurately capture the complex time-domain structure, thereby improving denoising performance. Secondly, our signal embedding method adeptly addresses the challenge of handling long temporal information in EEG signals. By intelligently grouping consecutive sampling points into patches, this method effectively preserves the context and temporal relationships of the signal, facilitating the network in learning and restoring features more efficiently. The sensitivity to long temporal information aligns with the characteristics of EEG signals, forming the basis for EEGDiR outstanding denoising effect. Therefore, the performance of EEGDiR in approaching noise-free signals arises not only from the superior processing of temporal information by the Retention mechanism but also from the mutually reinforcing capabilities of Retentive Network and signal embedding. This synergy enables the network to better comprehend and process the intricate structure of EEG signals.
\end{enumerate}

\section{Discussions}

In our analysis, we compared various denoising methodologies by examining their efficacy on the EOG, EMG, and SS2016 EOG datasets as presented in Table \ref{table3}. Our findings underscore the significant advancements our EEGDiR model, incorporating the Retentive Network and signal embedding techniques, makes over other methods.

\textbf{Comparative Analysis of Denoising Methods.}
The data reveals that simpler models like SCNN, despite their utility, fall short in more complex noise environments due to their basic structures and lack of advanced features like residual connections. Conversely, 1D-ResCNN improves upon this by utilizing diverse convolutional kernels and residual connections to mitigate overfitting, thereby enhancing denoising results. Notably, the LSTM model, which integrates temporal information directly into its structure, outperforms 1D-ResCNN by better handling the dynamics within EEG signals.
EEGDnet, leveraging the global modeling capabilities of the transformer architecture, enhanced by residual connections and layer normalization, significantly surpasses both SCNN and 1D-ResCNN. The most robust performance, however, is exhibited by EEGDiR. This model not only understands complex temporal sequences better due to the Retentive Network but also optimizes the handling of these sequences through our innovative signal embedding approach. This dual strategy is particularly effective in preserving the contextual and temporal integrity of EEG signals, which is crucial given their prolonged time-series nature.

\textbf{Performance Across Different Noise Levels.}
Our study also evaluated the performance of these methods across varying SNR levels, as depicted in Figures \ref{fig3}, \ref{fig4}, and \ref{new-1}. All methods demonstrated declining performance with decreasing SNR, highlighting the challenges posed by increased noise levels. However, EEGDiR consistently outperformed other methods at all noise levels, achieving the lowest RRMSE and highest correlation coefficients. This suggests that EEGDiR's architecture and signal processing strategies are well-suited to effectively reduce noise while preserving the integrity of the EEG signal.

\textbf{Broader Implications.}
The success of the Retentive Network in this context not only paves the way for its use in EEG signal denoising but also suggests its applicability to other types of temporal signals, such as electromagnetic and seismic data. The ability of the Retentive Network to process temporal information effectively, coupled with our signal embedding technique, offers a robust framework for denoising tasks across various domains requiring detailed temporal analysis. Moreover, this framework holds significant potential for enhancing downstream EEG tasks, such as classification \cite{lin2024eeg}, motor imagery \cite{gao2024multiscale}, brain recognition \cite{li2023eeg} and fatigue detection \cite{gao2023csf}, suggesting broad applicability in enhancing the accuracy and effectiveness of these complex applications.

\section{Conclusion}
By incorporating the Retentive Network architecture and employing signal embedding for processing EEG signals, this study introduces an innovative methodology aiming to leverage Retentive Network comprehensively for EEG signal denoising. The integration of Retentive Network architecture enhances the understanding and processing of temporal information in EEG signals, while the utilization of signal embedding underscores the processing of prolonged temporal information and feature extraction. Experimental results showcase the outstanding denoising performance of our proposed EEGDiR network on EOG, EMG  and SS2016 EOG datasets. In comparison to traditional EEG denoising methods, EEGDiR demonstrates notable enhancements in temporal information processing and global modeling.

The global modeling prowess of EEGDiR, coupled with its favorable handling of temporal information, positions it as an optimal choice for processing EEG signals. The incorporation of signal embedding further refines the representation of EEG signals, preserving context and temporal relationships more effectively. The synergistic application of the Retentive Network and signal embedding strategy yields a substantial improvement in the denoising performance of the EEGDiR network.

This study holds significant implications as a guide for integrating deep learning into neuroscience, offering valuable insights to enhance the efficacy and application potential of EEG signal processing. By providing an out-of-the-box deep learning dataset, our contribution enables subsequent researchers to expedite EEG signal denoising research by eliminating the need for extensive data preprocessing. This accelerates the development of EEG signal denoising methods.

\printcredits



\bio{}
\endbio

\endbio

\end{document}